\documentclass[aps,prb,twocolumn,superscriptaddress,nofootinbib,showpacs,floatfix]{revtex4-2}
\usepackage{comment}
\usepackage{physics}
\usepackage{amsmath,amsthm,amssymb,amsfonts, color, comment, graphicx}
\usepackage{dsfont}
\usepackage{xcolor}
\usepackage{float}
\usepackage{graphicx}
\usepackage{multirow}
\usepackage{dcolumn}
\usepackage{slashed}
\usepackage{bm}
\UseRawInputEncoding
\usepackage{hyperref}
\usepackage[T1]{fontenc}

\usepackage[normalem]{ulem}
\usepackage{physics}

\newcommand{\bea}{\begin{eqnarray}}
\newcommand{\eea}{\end{eqnarray}}

\newcommand{\bS}{\mathbf{S}}

\newcommand{\br}{\mathbf{r}}
\newcommand{\bM}{\mathbf{M}}

\newcommand{\ba}{\mathbf{a}}

\newcommand{\bn}{\mathbf{n}}

\newcommand{\be}{\begin{equation}}
\newcommand{\ee}{\end{equation}}

\newcommand{\bK}{{{\bf{K}}}}

\newcommand{\bQ}{{{\bf{Q}}}}

\newcommand{\beal}{\begin{align}}
\newcommand{\eeal}{\end{align}}

\newcommand{\upa}{\uparrow}
\newcommand{\dna}{\downarrow}

\def\l{\ell}

\newcommand{\btjstrw}{\mathrel{{\rotatebox[origin=c]{90}
{$\bowtie$}}\kern-0.18em\raisebox{-.95ex}{$\bullet$}
\kern-0.5em\raisebox{.97ex}{$\bullet$}
\kern-1.12em\raisebox{.97ex}{$\bullet$}
\kern-0.52em\raisebox{-.95ex}{$\bullet$}}}

\newcommand{\btjnbrR}{{\mathrel{\rotatebox[origin=c]{90}
{$\bowtie$}}\kern-0.22em\raisebox{.9ex}{$\bullet$}
\kern-1.em\raisebox{-.8ex}{$\bullet$}}}
\newcommand{\btjnbrL}{{\mathrel{\rotatebox[origin=c]{90}
{$\bowtie$}}\kern-0.22em\raisebox{-.8ex}{$\bullet$}
\kern-1.em\raisebox{+.9ex}{$\bullet$}}}

\def\a{\alpha}
\def\b{\beta}
\def\c{\chi}
\def\d{\delta}

\def\e{\epsilon}

\def\g{\gamma}

\def\l{\lambda}

\def\m{\mu}
\def\n{\nu}

\def\s{\sigma}

\def\ua{\uparrow}
\def\da{\downarrow}

\def\ml{{\mathcal{L}}}

\def\mo{{\mathcal{O}}}

\def\mr{{\mathcal{R}}}

\newcommand{\llangle}[1][]{\savebox{\@brx}{\(\m@th{#1\langle}\)}%
  \mathopen{\copy\@brx\kern-0.5\wd\@brx\usebox{\@brx}}}
\newcommand{\rrangle}[1][]{\savebox{\@brx}{\(\m@th{#1\rangle}\)}%
  \mathclose{\copy\@brx\kern-0.5\wd\@brx\usebox{\@brx}}}

\newcommand{\Z}{\mathbb{Z}}
\newcommand{\HH}{\mathcal{H}}
\newcommand{\Gsp}{G_{\text{space}}}

\begin{document}

\preprint{APS/123-QED}
\newcommand{\contribute}{$^{\parallel}$}
\newcommand{\correspond}{$^{\perp}$}
\renewcommand{\thefootnote}{}
\footnotetext{\contribute~These authors contributed equally to this work}
\footnotetext{\correspond~Corresponding author}

\title{Symmetry constrained field theories for chiral spin liquid to spin crystal transitions}
\author{Anjishnu Bose\contribute}
\email{anjishnu.bose@mail.utoronto.ca}
\affiliation{Department of Physics, University of Toronto, 60 St. George Street, Toronto, ON, M5S 1A7 Canada}
\author{Andrew Hardy
\contribute}
\email{andrew.hardy@mail.utoronto.ca}
\affiliation{Department of Physics, University of Toronto, 60 St. George Street, Toronto, ON, M5S 1A7 Canada}
\author{Naren Manjunath
\correspond\contribute}
\email{nmanjunath@perimeterinstitute.ca}
\affiliation{Perimeter Institute for Theoretical Physics, 31 Caroline St N, Waterloo, ON N2L 2Y5, Canada}
\author{Ramanjit Sohal}
\email{rsohal@uchicago.edu}
\affiliation{Pritzker School of Molecular Engineering, University of Chicago, Chicago, IL 60637, USA}
\author{Arun Paramekanti}
\email{arun.paramekanti@utoronto.ca}
\affiliation{Department of Physics, University of Toronto, 60 St. George Street, Toronto, ON, M5S 1A7 Canada}
\date{\today}
\setcounter{footnote}{0}
\renewcommand{\thefootnote}{\arabic{footnote}} 
\begin{abstract}
We consider the spin rotationally invariant Kalmeyer-Laughlin chiral spin liquid (CSL) in systems with broken time-reversal symmetry and explore symmetry constraints on possible conventional spin crystal states accessible via a direct transition. These constraints provide a framework to identify topological invariants of the magnetically ordered state. We show that the existence of a direct transition from a CSL requires a precise compatibility condition between the topological invariants of the ordered state and the anomaly of the CSL.
The lattice symmetries also constrain the functional form of the low-energy theory to describe these transitions. This allows us to construct explicit Chern-Simons-matter field theories for the transition into a class of noncoplanar orders identified as candidates directly accessible from the CSL, including the octahedral spin crystal on the kagom\'e lattice, and the tetrahedral order on the triangular and honeycomb lattice. These transitions can either be described using coupled fractionalized $\mathbb{CP}^1$ theories or fractionalized matrix principal chiral models. We also discuss extensions 
to more general magnetic ordering transitions out of the CSL. 
\end{abstract}

\maketitle

\section{Introduction}
Gauge theories coupled to matter fields offer a powerful framework for describing the physics of quantum spin liquid phases in frustrated magnets \cite{Balents2010, Zhou2017}. This language enables a deeper understanding of ground state properties of these topological phases, the emergent anyonic excitations, and the transitions into conventional broken symmetry phases. The fundamental interest in topologically ordered spin liquids and their transitions into magnetically ordered states \cite{Sachdev2011, Song2019, Song2020} stems from the fact that these lie beyond the Landau theory framework of conventional phases and  phase transitions \cite{Senthil2004b} and broaden our perspective of  gauge-matter theories \cite{Seiberg2016, Wang2017}. Such non-Landau transitions between topological order and symmetry broken order can also occur in fermionic systems such as transitions out of fractional quantum Hall liquids (FQH) \cite{Stormer1999}. The FQH state can arise through the `melting' of a charge Wigner crystal (CDW) \cite{Kivelson1987,Apalkov2010, Zhao2018, Villegas2021}, a concept that has recently received increased attention due to possible realizations in moir\'e platforms \cite{ Le2023,  Song2024, Haug2025}.\par

In this paper we consider the magnetic analogue of such a transition: a Kalmeyer-Laughlin (KL) chiral spin liquid (CSL) emerging from an ordered spin crystal with nonzero spin chirality. The CSL serves as a prototypical example of bosonic abelian topological order \cite{Kalmeyer1987, Laughlin1989, Marston1991}. This phase is described by a U$(1)_2$ Chern-Simons gauge theory and has an appealing analogy to a $\nu = 1/2$ fractional quantum hall state for bosons \cite{Kalmeyer1987, Laughlin1989}. This state can also be obtained by gauging a symmetry protected topological (SPT) phase of charge-$e$ bosons with Hall conductivity $\s_{xy}=2 e^2/h$ \cite{Senthil2013, Liu2013, He2015b, He2015c}. Recent numerical results have shown that topological or anyon 
superconductivity may emerge from an electronic CSLs upon doping with holes \cite{Wen1989_SC,Song2021, Divic2024b}.\par

The CSL was originally proposed to be the ground state of the triangular lattice Heisenberg model, which was later shown to have long-range coplanar $120^\circ$ antiferromagnetic order instead \cite{Huse1998, Li2022}. Nevertheless, the CSL remains a candidate for the ground state of the kagom\'e lattice Heisenberg model or its deformed variants \cite{Depenbrock2012,He2014,Kumar2014, Kumar2015, He2015}.
It also emerges as the ground state on the kagom\'e lattice with explicit chiral interactions which break parity and time-reversal symmetries \cite{Bauer2014, Wieteck2015, Wietek2017}.
Later work has found that such extended models on the square \cite{Sheng2024}, triangular \cite{Gong2017, Wietek2017, Hickey2017, Szasz2020, Cookmeyer2021}, and honeycomb models \cite{Hickey2016,Huang2021} also support regimes with CSL ground states.  Meanwhile, the equivalent of crystalline order in the context of spins that we consider in this work are noncoplanar magnetic states. Such states which exhibit nonzero spin chirality often border the KL-CSL in phase space \cite {Gong2015, Hickey2016, Gong2017, Wietek2017, Oliviero2022, Bose2023}. While experimental detection of both chiral spin liquids and noncoplanar order remains challenging, the CSL may occur in layered  materials \cite{Schweika2022} and  Herbertsmithite \cite{Herbertsmithite2018, HerbertsmithiteRevModPhys}. Noncoplanar phases have in turn been observed in breathing kagom\'e materials \cite{Gu2024} using X-ray scattering (RIXS) and Raman scattering \cite{Xiong2020,  Xiong2020, Biffin2024}. \par 

There are two complementary objectives of this paper. The general objective is to understand necessary conditions under which a given ordered state can possibly
be accessed from a CSL by tuning microscopic parameters. The subsequent specific goal is to construct a continuum field theory for the transition between a CSL and a particular class of noncoplanar orders that we define below. While the specific question can be  addressed using a field theory of bosonic spinons with a Chern-Simons term, we find it is easier to approach the general question from a more abstract perspective based on topological quantum field theory (TQFT). Accordingly, we organize the main arguments by developing the necessary background in Secs.~\ref{sec: RMO} and~\ref{sec:CSL-review}, addressing the general question in Sec~\ref{sec: compatibility}, and writing the field theory in Sec.~\ref{sec: field theory}.\par

The general condition we propose is motivated by the fact that different lattices with the same symmetry give rise to topologically distinct CSLs. For instance, the CSLs on the kagom\'e and triangular lattice have the same symmetry group $G = \mathrm{SO}(3) \times p6$, but are topologically distinct because they have different \textit{anomalies} of the $G$ symmetry, meaning that the spin-$1/2$ anyons in the CSL localize on different sites in the real-space unit cell in each case. Here we interpret the anomaly as a Lieb-Shultz-Mattis constraint \cite{Hastings2003LSM,Po2017LH,Cho2017LSM,Ye2022LSM}; we review these notions for the CSL case in Sec.\ref{sec:CSL-review}. This observation motivates the question of whether ordered states on different lattices should also be considered topologically distinct, even if they have the same symmetry group. 

In Sec.\ref{sec: RMO}, we propose a topological description of the quantum ordered state that incorporates symmetry as well as a topological invariant that contains lattice information; in this sense it goes beyond the classical results of~\cite{Messio2011}.
We do so by viewing the ordered state as a \textit{parameterized family} of gapped ground states \cite{Else2021,Wen2023,Manjunath2024}; the connection between parameterized families and magnetic orders has not appeared before to our knowledge.
This invariant can in principle be characterized by introducing a topological defect of the order parameter and measuring an associated response such as a Berry phase or some generalization of it. 
Then in Sec.~\ref{sec: compatibility}, we use a `compatibility relation' recently proposed in Ref.~\cite{Manjunath2024} to obtain a necessary condition for a continuous transition. This condition relates the anomaly of the CSL to the above topological invariant of the ordered state. Failing the compatibility condition implies that the Hilbert spaces of the spin liquid and the ordered state are different, e.g. one phase has spin-1/2 degrees of freedom at certain points on the lattice while another phase has integer spins at those sites. In this scenario, it is not even natural for the two phases to be in the same phase diagram, let alone to have a continuous transition between them. 
In the process, we clarify some subtleties in the prior literature on crystal symmetry fractionalization in CSLs and its relation to the anomaly of the CSL \cite{Zalatel2016, Cincio2015, Song2023}. 
\par

We report our next main result in Sec.\ref{sec: field theory} where we construct a continuum field theory describing the transition between the topological CSL and a restricted class of symmetry broken noncoplanar orders called regular orders \cite{Messio2011}. 
We focus on systems with \(\mathrm{SO}(3)\times p6\) symmetry and explicit time-reversal symmetry (TRS) breaking.
This symmetry restriction turns out to exclude precisely the noncoplanar proximate states, the cuboctahedral orders (cuboc-1 and cuboc-2), 
that exhibit first-order transitions into the CSL \cite{Gong2015}, although it includes specific noncoplanar orders as the tetrahedral and octahedral states.

For the orders considered here, we use the machinery of Chern-Simons-matter field theories in two equivalent descriptions : (a) coupled multi-flavor \(O(3)\) nonlinear sigma-models (NL\(\s\)M) with a fractionalized \(\mathbb{CP}^1\) formulation, (b) \(\mathrm{SO}(3)\) NL\(\s\)Ms and a novel fractionalization using gauged-\(\mathrm{U}(2)\) matrix partons. We conclude by discussing possible numerical checks on these theories in Section \ref{sec: Discussion} .\par
\section{Regular Magnetic Orders} \label{sec: RMO}

\subsection{Review: Classical definition}

Let us review the notion of regular magnetic orders (RMOs) and understand the different types of RMOs that occur on various lattices. Ref.~\cite{Messio2011} defined an RMO as a classical spin configuration which breaks the global SO$(3)$ spin rotation symmetry down to a subgroup but preserves the space group symmetry $\Gsp$ of the lattice. 
However, certain space group symmetry operators may need to be modified by SO$(3)$ spin rotations.
A magnetic order is not regular if some space group operation cannot be made a symmetry even after multiplying by spin rotations. Using group-theoretic arguments, Ref.~\cite{Messio2011} classified all the classical RMOs on the triangular, honeycomb, and kagom\'e lattice. 

We can in principle include anti-unitary operations such as reflections or time-reversal in the definition of $\Gsp$; in the latter case, we would refer to $\Gsp$ as a `magnetic space group'. However, in this paper we only consider RMOs with broken time-reversal and with broken reflection symmetry. 
This restricts $\Gsp$ to one of the 5 orientation-preserving wallpaper groups $p1, p2, p3, p4$ or $p6$.  This is relevant to the cases where there is an explicit chiral term in the Hamiltonian which breaks reflection and TRS. 
These restricted orders are enumerated in Table \ref{table: Noncoplanar RMOs}, Table \ref{table: coplanar and collinear RMOs}, and Table \ref{table: ferro collinear RMOs}. In particular, the noncoplanar cuboc-1 and cuboc-2 orders on the kagom\'e lattice are not regular for broken time-reversal systems, and are therefore not covered by the analysis in this paper. \par

\begin{figure}[!ht]
 \centering
    \includegraphics[width=0.35\textwidth]{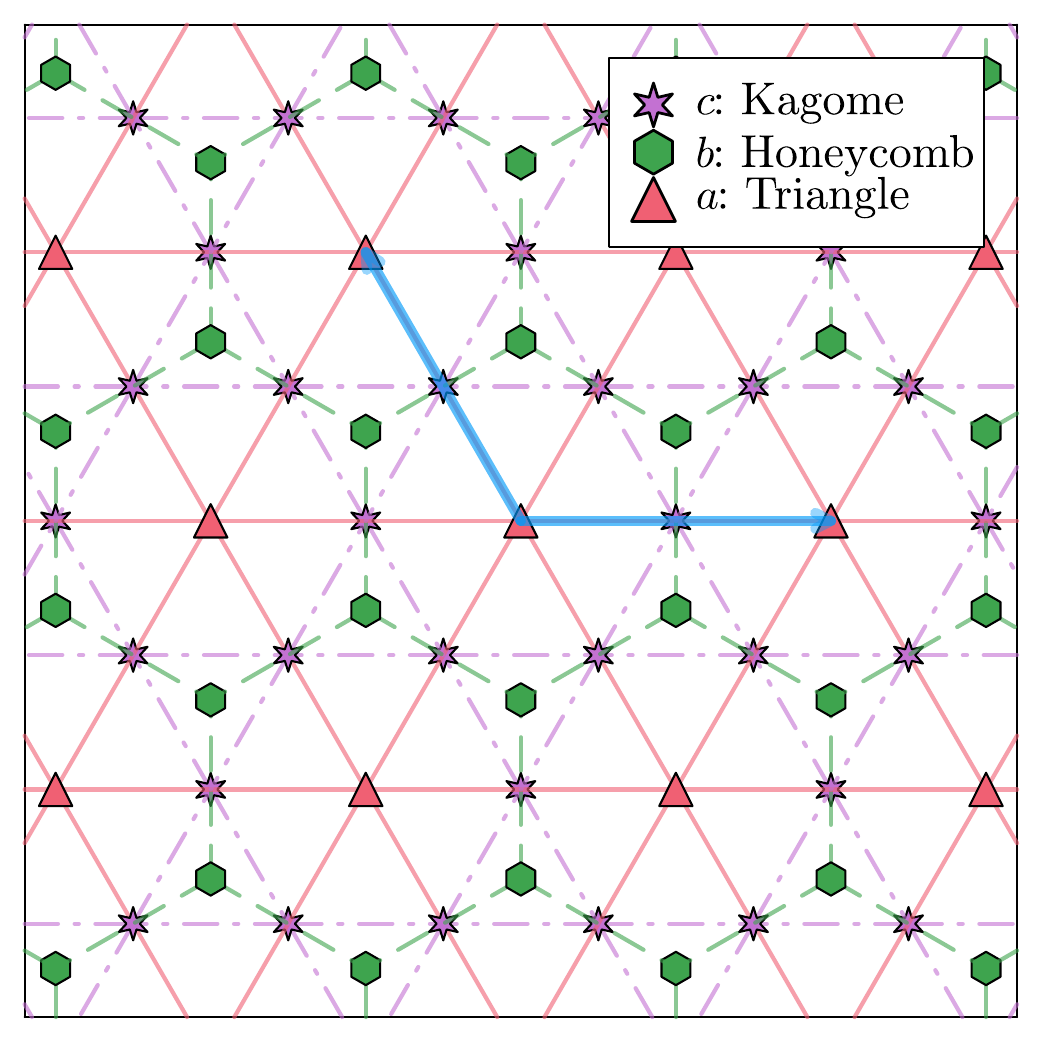}
 \caption{The three high-symmetry points \(a, b,\) and \(c\) on a lattice with \(p6\) spatial symmetry consisting of two translations \(T_{\ba_1}\) and \(T_{\ba_2}\) (marked in blue). The unit cell consists of one \(a\) site, two \(b\) sites, and three \(c\) sites. The \(a\) sites are \(C_6\) centers, \(b\) sites are \(C_3\) centers, and \(c\) sites are \(C_2\) centers. On a triangular lattice, \(a\) sites are the vertices, \(b\) sites are the plaquette centers, and \(c\) sites are the edge centers. On a honeycomb lattice, \(a\) sites are the plaquette centers, \(b\) sites are the vertices, and \(c\) sites are the edge centers. On a kagom\'e lattice, \(a\) sites are the hexagonal plaquette centers, \(b\) sites are triangular plaquette centers, and \(c\) sites are the vertices. }
\label{fig:p6 lattices}
\end{figure}

The same symmetry group $\Gsp$ can be realized on different lattices. For instance, the triangular, honeycomb and kagom\'e lattices can all have $p6$ space group symmetry. The difference between them lies in the placement of the physical orbitals at different high-symmetry points in the real-space unit cell, as illustrated in Fig.\ref{fig:p6 lattices}. For the triangular lattice, the physical orbitals are located on the $C_6$ symmetric points, for the honeycomb lattice they are located at the $C_3$ symmetric points, and for the kagom\'e lattice, the physical orbitals are located at the $C_2$ symmetric points. \par 

The above classical definition of an RMO encodes its symmetry in a precise fashion but does not specify possible topological invariants of the quantum ordered state. This makes it difficult to compare the properties of a particular ordered state with those of a neighboring spin liquid, in which both the symmetry and various topological invariants (such as the number of anyons, their fusion and braiding, or their symmetry quantum numbers) can be captured. It will be useful to define RMOs in a way that specifies both their symmetries and any topological invariants they may carry. This can be done if we define an RMO as a \textit{parameterized family} of gapped ground states.

The topological terms we consider can often be understood as Berry phase terms in a nonlinear sigma model for the order parameter. When they do not affect the low-energy properties in the ordered state, they are often not included in the low-energy description. We explicitly keep them in our definitions because this allows us to formulate a `compatibility relation' between a topological invariant of the CSL (specifically the anomaly of the CSL, defined in Sec.\ref{sec:anomaly}) and the topological invariants of the ordered state. This compatibility relation provides a necessary (but not sufficient) condition for a continuous transition to be possible between a topologically ordered state such as a CSL and a symmetry-broken state. We will discuss the compatibility relation in Sec.\ref{sec: compatibility}. In the rest of this section, we give the general definition of a parameterized family, and explain how it captures other RMOs (including noncoplanar orders on the triangular or kagom\'e lattices).

\subsection{Warmup: Ferromagnetic order as a parameterized family}

Before giving the general definition of a parameterized family, let us illustrate the main idea with the example of a ferromagnet. This is a simple ordered state in which the SO$(3)$ spin rotation symmetry is broken down to an SO$(2)$ subgroup (spin conservation along an axis). The classical definition of the ferromagnet along the lines of Ref.~\cite{Messio2011} is that the full symmetry of the ordered state is SO$(2) \times \Gsp$, and the space group operations in the ordered state are given by the `bare' spatial transformations without any accompanying SO$(3)$ operation (see Table \ref{table: ferro collinear RMOs}). The `ferromagnetic state' in the quantum setting is in fact one member of a \textit{family} of gapped ground states with symmetry SO$(2) \times \Gsp$, each being specified by a particular choice of magnetization axis.
Different states in the family are related by a global SO$(3)$ rotation. The parameter space for this family is the 2-sphere $S^2$, which is also the space of all possible magnetic field directions. $S^2 = \text{SO}(3)/\text{SO}(2)$, i.e. $S^2$ parameterizes the different ways in which SO$(3)$ can be broken down to SO$(2)$. \par

We can associate the family of ferromagnetic states with a topological invariant which is due to the topology of $S^2$, and is not obvious from a single state. To see this, first consider a single spin in a magnetic field, where the different choices of $\vec{B}$ also give a family of states over $S^2$. The topological invariant associated to this zero-dimensional family is the Chern number $C$ of the Berry connection over the space $S^2$ defined by
\begin{equation}
    C  = \frac{1}{2\pi} \int_{S^2} (\partial_{B_x}\bra{\psi} \partial_{B_y} \ket{\psi}-\partial_{B_y}\bra{\psi} \partial_{B_x} \ket{\psi}), 
\end{equation}
where $\ket{\psi}$ is the ground state for a particular choice of magnetic field. We call this the Berry-Chern number. In a field-theoretic description of the spin, the nonzero Chern number would be fixed by a quantized Berry phase term for the $S^2$ order parameter. One can show that for a spin-$j$ system with ground state carrying $S_z$ eigenvalue $m$ (where the $z$ axis is defined to be the magnetic field direction), the Berry-Chern number is $C=2m$. This is a topological invariant since it is integer-quantized, and $C$ cannot be changed unless we perform a process in which the magnitude of the magnetic field becomes zero somewhere. Such a process would close the gap for at least one ground state in the family.

For a two-dimensional ferromagnet with lattice symmetry $\Gsp$, we can place a certain number $C_{\alpha}$ of ferromagnetic spins at each high symmetry point $\alpha$ in the real-space unit cell. 
The ferromagnet will then be specified by an assignment of symmetry-resolved Berry-Chern numbers $\{C_{\alpha}\}$, although two assignments related by the space group symmetry are not unique. 
For instance, let $\Gsp = p6$, and  $a,b,c$ denote the triangular, honeycomb and kagom\'e sites respectively, as in Fig.~\ref{fig:p6 lattices}. 
Each configuration is determined by some triple $(C_a, C_b,C_c) \in \Z^3$ depending on the total $S_z$ eigenvalue at each site. 
The simplest spin-$1/2$ ferromagnet on the kagom\'e lattice would satisfy $C_a = C_b = 0, C_c = 1$, but different triples could be rendered topologically equivalent through deformations which respect $p6$ symmetry. For example we can imagine adiabatically moving 6 total spin-$1/2$ objects from each $a$ site to the $b$ sites while preserving $p6$ symmetry, without closing any gap. This reduces $C_a$ by $6$ but increases $C_b$ by 3, since each $b$ site receives one degree of freedom from its three neighboring unit cells. This leads to the equivalence
\begin{equation}
    (C_a, C_b, C_c)\simeq (C_a-6, C_b + 3, C_c).
\end{equation}
If we consider deformations that move spins from $a$ to $c$, we find a second equivalence
\begin{equation}
    (C_a, C_b, C_c) \simeq (C_a-6, C_b, C_c+2).
\end{equation}
To derive the full classification, we need to consider the naive $\Z^3$ classification modulo these equivalences. 
The integer $C_{tot} := C_a + 2 C_b + 3 C_c$ is invariant under both the equivalences. A second invariant is $C_a \mod 6$. 
Therefore the full classification is given by the group $\Z \times \Z_6$.
A formal mathematical scheme to classify parameterized families has been discussed previously in \cite{Manjunath2024}, and we review this briefly in Appendix~\ref{app:families}. 
Those methods show that $\Z \times \Z_6$ is indeed the full classification of families over $S^2$ with a common $p6$ space group symmetry in each state. Instead of the invariants $C_{tot}, C_a \mod 6$ one could also work with the three invariants $C_{tot},C_b \mod 3, C_c \mod 2$, and this would give an equivalent parameterization \footnote{The above classification of families is exactly the same as the classification of interacting topological crystalline insulators with symmetry $\mathrm{U}(1) \times \Gsp$, if we consider only the topological terms that contribute to the charge response, and set the Hall conductance to zero. 
Such terms assign integer electric charges to different high symmetry points, and are classified e.g. in \cite{Manjunath2021,Song2020RS} (for free fermions, see \cite{Kruthoff2017bandCombinatorics}).
When classifying ferromagnets we are interested in the same problem, just with replacing electric charge by the Chern number of a magnetized spin. }.

To summarize:
\begin{enumerate}
    \item The classical definition of the ferromagnet is that it is a spin configuration with symmetry $SO(2) \times \Gsp$, where $\Gsp$ consists of bare space group operations.
    \item We prefer to define the ferromagnet as a parameterized family of states. This definition involves two levels of data. First we have symmetry data, including the symmetries of the disordered state ($G = \mathrm{SO}(3) \times \Gsp$) and of each ordered state ($H = SO(2) \times \Gsp$). From this we find that the parameter space for the family is $\Lambda\equiv G/H = S^2$. Next we have a topological invariant (the Chern numbers $C_a, C_b, C_c$ modulo some equivalences) whose classification is specified by $\Lambda$ and $\Gsp$. 

\end{enumerate}

\subsection{General definition of parameterized families}

A general parameterized family is similarly given by two levels of data which we unpack below:
\begin{enumerate}
    \item The symmetry data, specified by four elements $(G,H,\Lambda = G/H, H_0)$; 
    \item A topological invariant which depends on $\Lambda$ and $H_0$.
\end{enumerate}

$G$ is the symmetry of the disordered state; here we always assume $G = \mathrm{SO}(3) \times \Gsp$. $H$ is the symmetry of any given ordered state.  We can define an order parameter for this symmetry breaking which takes values in a space $\Lambda= G/H$. 
A \textit{parameterized family} of gapped ground states is a set of ground states $\{ \ket{\psi_{\lambda}} \}$, one for each value of the order parameter $\lambda \in \Lambda$.
These states are all related by a $G$ action. $H$ is a stabilizer of $G$, which means $H \subset G$ and consists of the elements that do not modify the spin configuration of a specific choice $\ket{\psi_{\lambda}}$ of the ordered state.

Additionally we specify a group $H_0$, which is the \textit{common} symmetry of each state in the family. Mathematically, $H_0$ is the largest subgroup of $H$ that remains invariant under conjugation by $G$. It is therefore a normal subgroup of $G$ (and $H$). \par

The RMO portion (1) of paramaterized families can be thought of in analogy to projected symmetry group (PSG) classifications  \cite{Wen2002, Messio2011}. 
In this analogy, $G$ is the ungauged equivalent of $\Phi$, the local gauge transformations of the PSG. $H$ corresponds to the residual symmetry or PSG, and the subgroup $H_{spin} \subset \mathrm{SO}(3) \cap  H$ corresponds the internal gauge group, or IGG. 
\par
In the case of RMOs, $H_0$  is comprised of operations in $H$ which do not include any spin rotations in SO$(3)$. (These are the only operations that would remain in $H_0$ when conjugated by infinitesimal SO$(3)$ rotations.) For the ferromagnet, $H_0 = \Gsp$ since the definition of each crystalline symmetry operator does not include any spin rotations. Note that $H_0$ cannot contain any subgroup of SO$(3)$. 
Even though each state in the ferromagnet preserves an SO$(2)$ subgroup, this subgroup is generated by different operators $e^{i \frac{\theta}{2} \hat{n} \cdot \vec{\s}}$ for different choices of magnetization axis $\hat{n}$.
Therefore, no subgroup of the spin rotation symmetry SO$(3)$ is common to all the states in the family. In general, $H_0$ could further reduce to some subgroup of $\Gsp$, if crystalline elements are also modified by rotations in SO$(3)$.

\begin{table*}[!ht]
\centering
\begin{tabular}{ |c|c|c|c|c|c|c|c| } 
     \hline
     Order & Lattice & \(\tilde{T}_{\ba_1}\) & \(\tilde{T}_{\ba_2}\) & \(\tilde{C}_{3, a}\) & \(\tilde{C}_{3, b}\) & \(\tilde{C}_{2, a}\) & \(\tilde{C}_{2, c}\)\\
     \hline 
     Octahedral & kagom\'e & \(\mr_{\pi}(\hat{y})\times T_{\ba_1}\) & \(\mr_{\pi}(\hat{x})\times T_{\ba_2}\) & \(\mr_{2\pi/3}(\hat{n}_{-})\times C_{3, a}\) & \(\mr_{2\pi/3}(\hat{n}_{+})\times C_{3, b}\) & \(\mathds{1}\times C_{2, a}\) & \(\mr_{\pi}(\hat{z})\times C_{2, c}\) \\
     \hline 
     Tetrahedral & Triangular & \(\mr_{\pi}(\hat{y})\times T_{\ba_1}\) & \(\mr_{\pi}(\hat{x})\times T_{\ba_2}\) & \(\mr_{2\pi/3}(\hat{n}_{-})\times C_{3, a}\) & \(\mr_{2\pi/3}(\hat{n}_{+})\times C_{3, b}\) & \(\mathds{1}\times C_{2, a}\) & \(\mr_{\pi}(\hat{z})\times C_{2, c}\) \\
     \hline 
     Tetrahedral & Honeycomb & \(\mr_{\pi}(\hat{y})\times T_{\ba_1}\) & \(\mr_{\pi}(\hat{x})\times T_{\ba_2}\) & \(\mr_{2\pi/3}(\hat{n}_{-})\times C_{3, a}\) & \(\mr_{2\pi/3}(\hat{n}_{+})\times C_{3, b}\) & \(\mathds{1}\times C_{2, a}\) & \(\mr_{\pi}(\hat{z})\times C_{2, c}\) \\
     \hline 
\end{tabular}
\caption{The modified residual lattice symmetries of all regular anti-ferromagnetic noncoplanar orders in systems with \(G=p6\times \mathrm{SO}(3)\) symmetry where \(p6\) includes translations \(T_{\ba_1}\,,T_{\ba_2}\) and rotations \(C_{6, a}\). For convenience, one might wish to break the \(C_6\) into a \(C_3\) and an inversion \(C_2\), which may also be on different centers \(C_{3, b}=T_{\ba_1}^{-1}\cdot C_{3, a}\) and \(C_{2, c}=C_{2, a}\cdot T_{\ba_1}\). Each of the orders have an enlarged \(2\times 2\) magnetic unit cell and are described by three characteristic wave-vectors \(\bQ = \bM_{i}\), \(i\in\{1, 2, 3\}\). 
Time-reversal and reflection are assumed to be broken explicitly which restricts the set of possible orders to a subset of those reported in \cite{Messio2011} eg. cuboc orders on the kagom\'e lattice and cubic order on the honeycomb lattice are not regular when time-reversal is absent.
Each modified lattice symmetry has an attached spin rotation \(\mr_{\theta}(\hat{n})\) by an angle \(\theta\) around a direction \(\hat{n}\) on the Bloch sphere. \(\hat{n}_{+}=(1/\sqrt{3})(1, 1, 1)\) and \(\hat{n}_{-}=(1/\sqrt{3})(-1, 1, 1)\). 
It is important to note that the exact axes mentioned in the table are all modulo a global spin rotation ; all the physics is dictated by relative spin-rotations between the different symmetries which are captured by operators like \(\tilde{T}_{\ba_1}\cdot \tilde{T}_{\ba_2}\cdot \tilde{T}_{\ba_1}^{-1}\cdot \tilde{T}_{\ba_2}^{-1}\) and \(\tilde{C}_6^6\) being \(\pm \mathds{1}\). 
All these RMOs will have a \(\mathds{Z}_2\) topological invariant classifying their families. See Table \ref{table: coplanar and collinear RMOs} and Table \ref{table: ferro collinear RMOs} for simpler coplanar and collinear orderings.}
\label{table: Noncoplanar RMOs}
\end{table*}

\subsection{Example: Octahedral order}

Now let us consider the octahedral order on the kagom\'e lattice. Assuming $\Gsp = p6$, the symmetry data is given by $G = \mathrm{SO}(3) \times p6$ and $H = p6$.\

\begin{figure}[!ht]
 \centering
    \includegraphics[width=0.35\textwidth]{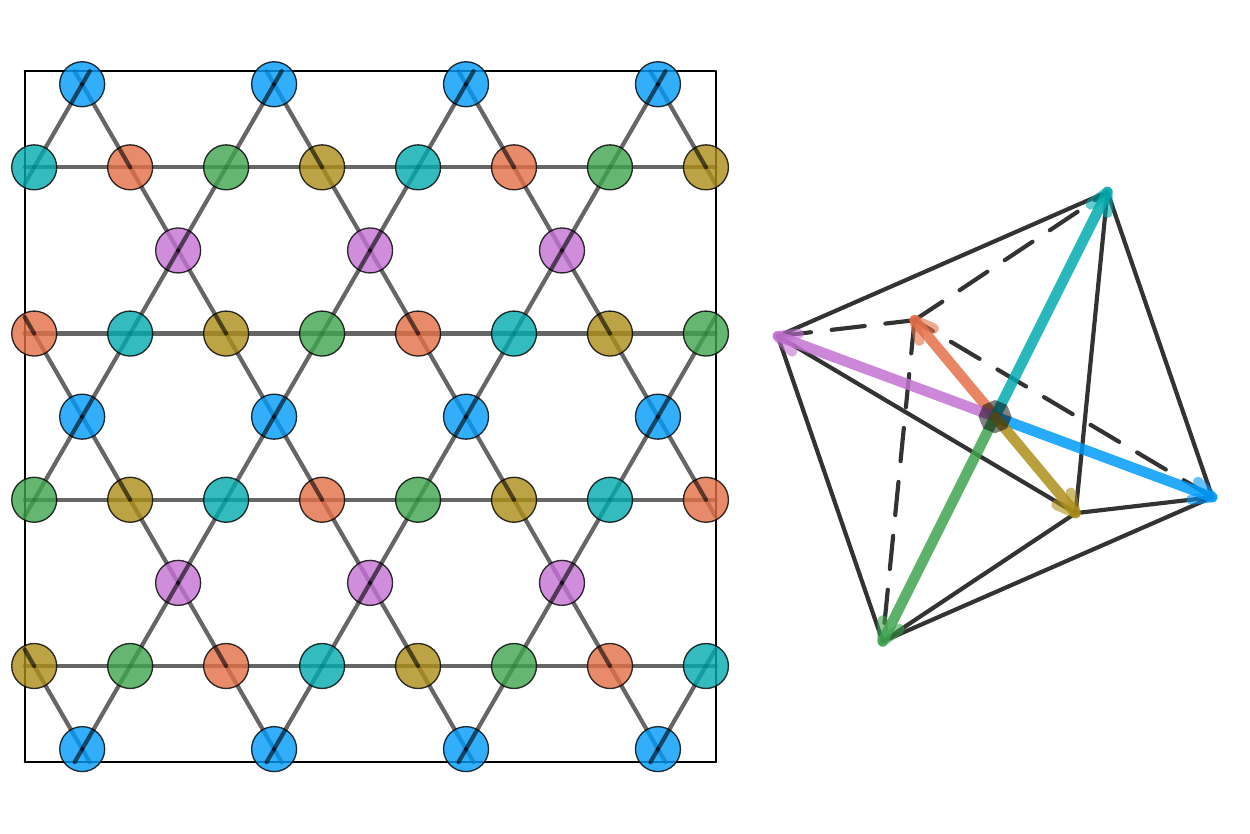}
 \caption{The octahedral order on the kagom\'e lattice with the spins pointing along the six vertices of an octahedron and spanning a \(2\times 2\) unit cell with \(12\) sites. 
 Each of the three sublattices exhibits a N\'eel order at the three \(\bM\) points in the Brillouin zone. The triangular plaquettes host a uniform scalar spin chirality.}
\label{fig:p6 Octahedral orders}
\end{figure}

The translation generators in $H$ are modified by $\pi$ rotations about orthogonal spin axes, while the rotation generator is also modified by a $2\pi/3$ rotation about the \((1,1,1)\) axis (see Table \ref{table: Noncoplanar RMOs} and Fig. \ref{fig:p6 Octahedral orders}). 
Therefore, the group $H_0$ is a subgroup of $p6$ generated by even-order translations and by the $C_2$ rotation, which is invariant upon conjugation by $SO(3)$. 
We will write $H_0 = \tilde{p2}$, with the tilde denoting that $H_0$ is isomorphic to the group $p2$ but with a $2\times 2$ unit cell. Finally, the parameter space $\Lambda = G/H$ is equivalent to SO$(3)$ as a topological space, and is referred to as the real projective space $\mathbb{RP}^3$. 

In zero dimensions, parameterized families over $\mathbb{RP}^3$ can be realized by three spins pointing along different axes, and are $\Z_2$ classified. Since the fundamental group of $\mathbb{RP}^3$ is $\pi_1(\mathbb{RP}^3) = \Z_2$, it is possible to define a $\Z_2$ invariant corresponding to the Berry phase 0 or $\pi$ acquired by a state upon being transported through a nontrivial cycle in $\mathbb{RP}^3$. (The precise connection between homotopy groups and the classification of parameterized families is discussed in Appendix~\ref{app:families}.)
Such a cycle can be implemented by a $2\pi$ rotation of the order parameter about some axis. A $\pi$ Berry phase can only be realized if the ordered state overall transforms as a half-integer spin object under SO$(3)$.

For the two-dimensional case with spatial symmetry $H_0 = \tilde{p2}$, the classification is obtained by enumerating the different tilings of this order parameter within the $2\times 2$ unit cell that respect the $\tilde{p2}$ symmetry. 
Since there are four inequivalent sites in the $\tilde{p2}$ unit cell, we have an overall $\Z_2^4$ classification. The octahedral order on the kagom\'e lattice is one specific element within this classification, and is realized by placing ordered degrees of freedom only on those sites of the $\tilde{p2}$ lattice which correspond to the original kagom\'e sites. 

\subsection{Generalization to other noncoplanar orders}

\begin{figure}[!ht]
 \centering
    \includegraphics[width=0.5\textwidth]{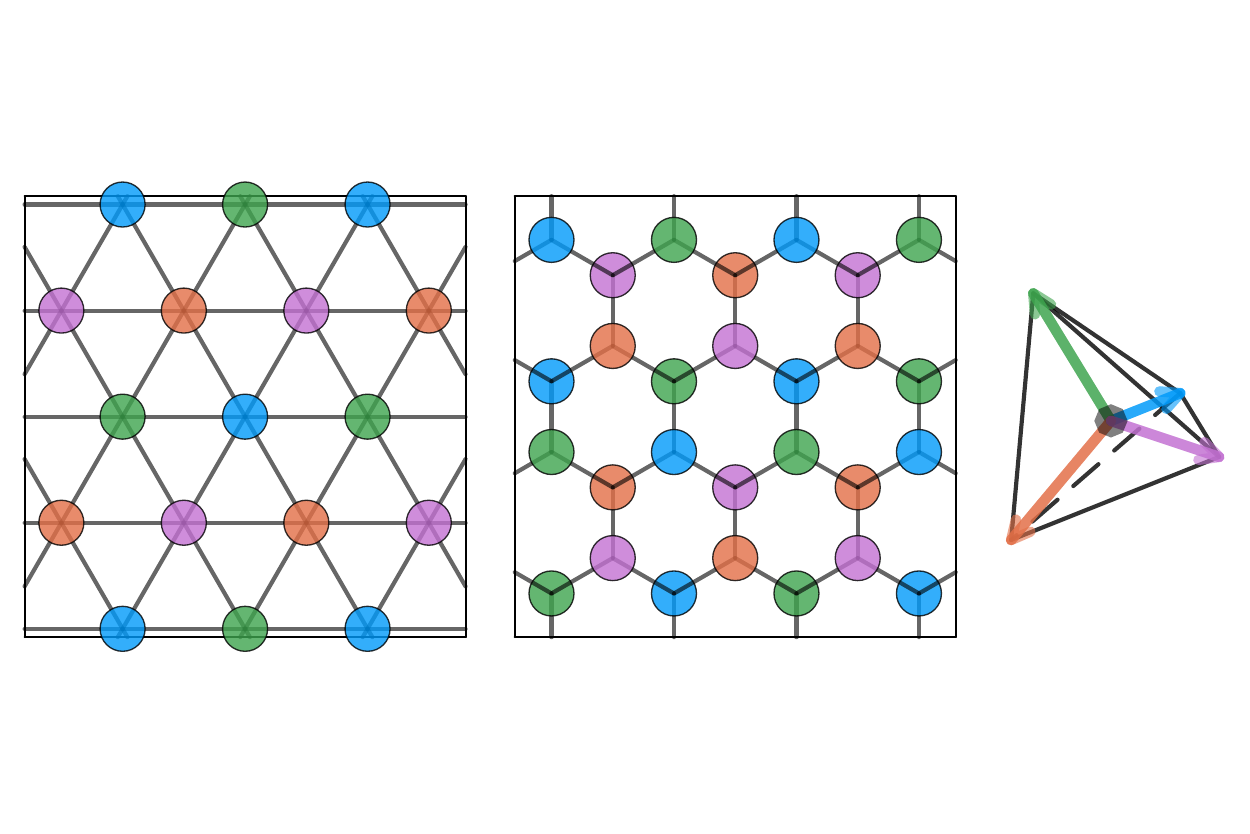}
 \caption{The tetrahedral order on the triangular (Left) and honeycomb (Right) lattices with the spins pointing along the four vertices of a tetrahedron spanning a \(2\times 2\) unit cell with \(4\) and \(8\) sites respectively. 
 The tetrahedral order can be obtained by overlapping three N\'eel orders at the three \(\bM\) points in the Brillouin zone (see Table \ref{table: Noncoplanar RMOs expectations}). It exhibits uniform scalar spin chirality just like the octahedral order.}
\label{fig:p6 Tetrahedral orders}
\end{figure}

The above reasoning can be applied to noncoplanar orders on different lattices. For instance, consider the tetrahedral order on the honeycomb or the triangular lattice as illustrated in Fig. \ref{fig:p6 Tetrahedral orders}. Similar to the octahedral order, these orders have symmetry groups which satisfy (see Table \ref{table: Noncoplanar RMOs})
\begin{align}
    \tilde{T}_{\bf a_1} \tilde{T}_{\bf a_2} \tilde{T}_{\bf a_1}^{-1} \tilde{T}_{\bf a_2}^{-1} = -\mathds{1}\,, \nonumber \\
    \tilde{C}_6^6 = +\mathds{1}\,.
\end{align}
Here $\tilde{T}_{\bf a_i},\tilde{C}_6$ denote the space group symmetries of the tetrahedral order after modification by spin rotations. The group structure for this order is the same as for the octahedral order on the kagom\'e lattice.
Therefore, the data $G,H,H_0,\Lambda$ are the same in each case. Equivalently, they are all triple-$Q$ orders at the $M$ point.
However, they are captured by different topological invariants, assuming the ordered variables all transform as spin-$1/2$ objects. The tetrahedral order on the triangular lattice would correspond to a family over $\mathbb{RP}^3$ tiled over a $\tilde{p2}$ unit cell, but only on the points corresponding to triangular sites.

\section{Symmetry fractionalization and anomaly of the kagom\'e Chiral spin liquid}\label{sec:CSL-review}

In this section, we review the concepts of topological order, symmetry fractionalization and anomaly, and the arguments explaning why there is a unique CSL on the kagom\'e lattice.
We also clarify a point of potential confusion in the prior literature on symmetry fractionalization in the CSL.

\subsection{Topological order}
We say that a gapped two-dimensional system has \textit{topological order} if it has a ground state degeneracy on a torus, with excitations that are anyonic, i.e. have fractional statistics. The CSL is a topologically ordered state with precisely one nontrivial anyon (the semion), in addition to the trivial boson. 
Topological order with only abelian anyons can be fully captured by a $K$ matrix Chern-Simons theory.  
The entire content of the topological order, including the fusion and braiding of each anyon, is specified by the $K$ matrix alone. 
Thus the $K$ matrix for the Kalmeyer-Laughlin CSL is identical to that of the bosonic spin-$1/2$ Laughlin state; the difference between the two lies in the additional symmetries assumed in each case \cite{Senthil2013, He2015b}. \par 

The relevant properties of the \(K\)-matrix which characterizes the KL-CSL phase are : (a) \(|\det(K)|=2\) which encodes the double degeneracy on a torus and the existence of two anyons, the trivial boson and the semion, (b) \(|sgn(K)|=1\) (difference in number of positive and negative eigenvalues) which encodes the existence of a single chiral edge mode, and (c) the diagonal entries must be even integers in a bosonic system. The simplest \(K\)-matrix which satisfies this is obviously \(K=2\), but in general can have larger matrix solutions as well which will be used in Sec.\ref{sec: field theory}. Also note that \(K\)-matrices are not unique as their topological properties stay invariant under a \(GL(\Z)\) transformation \cite{ Lu2012, Lu2016, Benini2018, Shyta2022}.

\subsection{Symmetry fractionalization}
Suppose the topologically ordered state has additional global symmetry, given by the group $G$.  \textit{Symmetry fractionalization} is the phenomenon in which the anyons acquire fractional quantum numbers under $G$.
In the $K$ matrix formulation, the symmetry fractionalization can be encoded in different `charge vectors' that couple the topological order to various background gauge fields for the $G$ symmetry:
\begin{equation}\label{eq:SF-Kmatrix}
    \mathcal{L}_{frac} = \frac{q_I}{2\pi} A \wedge da^I + \dots 
\end{equation}
where $A$ denotes a background $G$ gauge field, and the dynamical gauge fields $a^I, I = 1,2, \dots$ specify the topological order. In our case, where the symmetries are unitary and do not permute anyons in the CSL, the charge vector completely specifies the symmetry fractionalization. The charge vector is particularly well-known in the context of fractional quantum Hall states, which have $\mathrm{U}(1)$ charge conservation symmetry. 
In that context, $A$ can be interpreted as a magnetic vector potential, and the above term assigns a fractional charge to each anyon. In the $K$ matrix language, the charge of the anyon represented by the integer vector $\vec{b}$ is
\begin{equation}
    Q_b = \vec{b}^T K^{-1} \vec{q} \mod 1.
\end{equation}
The choice of $\vec{q}$ and $Q_b$ is not completely arbitrary. For instance, it must obey the anyon fusion rules:
\begin{equation}\label{eq:SF-condition}
    Q_{b_1} + Q_{b_2} = Q_{b_1 \times b_2} \mod 1,
\end{equation}
where $b_1 \times b_2$ is the fusion product of $b_1, b_2$. (This constraint can also be obtained by demanding gauge invariance of the $A da$ term.) Since two semions fuse to the identity, the semion can only carry 0 or $1/2$ fractional quantum numbers under a given symmetry. Moreover, there can be multiple independent quantum numbers $Q^{(1)}_b,Q^{(2)}_b,\dots$ depending on the number of independent symmetry generators.

Above, we could take $G$ to be an internal symmetry, a spatial symmetry, or a combination of the two. Let us now specialize to $G = \mathrm{SO}(3) \times p6$ and find the number of distinct quantities $Q^{(j)}_s$. First, the semions can carry either spin 0 or spin-$1/2$ under SO$(3)$. 
Therefore we can label $Q^{(1)}_s$ as the semion spin. Note that in the usual parton constructions of CSLs, we always set $Q^{(1)}_s = 1/2$; the other case is mathematically allowed but less natural to obtain from a parton construction. 

Next we turn to the crystal symmetry fractionalization. 
Apart from reviewing some definitions below, we also wish to clarify a subtlety that arises when comparing the TQFT description used in Ref.~\cite{Ye2024} (which generalizes the above $K$ matrix description) to the more direct approaches used in Refs.~\cite{Zalatel2016,Cincio2015,Song2023}.

In the TQFT approach, the abstract way to compute symmetry fractionalization data $Q^{(j)}_s$ uses the following setup.
We imagine creating two semions and separating them far apart. 
Then, we perform a sequence of group operations in some neighborhood around one semion, which sum to the identity.
Symmetry fractionalization of the semion means that under these operations the state can acquire a complex phase $e^{2\pi i Q_s^{j}}$. 
Here it is important to note that the usual TQFT description is only mathematically well-defined for internal symmetries. 
To include spatial symmetries, one needs to make the assumption (known as the `crystalline equivalence principle' \cite{thorngren2018,manjunath2022mzm}) that spatial symmetries act effectively as internal symmetries on the low-energy theory. Therefore, we should consider the symmetry of the TQFT to be a group $G = SO(3) \times \Gsp$ that effectively acts as an internal symmetry.
The crystalline equivalence principle has been checked in detail for bosonic systems without topological order by matching its predictions to classifications obtained by alternative methods \cite{Huang2017,Song2020RS}.
Like Ref.~\cite{Ye2024}, we will assume that it also holds for topologically ordered states and try to understand the consequences.

When $\Gsp$ is orientation-preserving, the number of independent invariants $j = 1,2, \dots$ is equal to the number of distinct high-symmetry points in the real-space unit cell. For $p6$ symmetry, it can be shown that a complete set of independent operations is
\begin{align}
    e^{2\pi i Q^{(2)}_s} &= \arg \bra{\psi_s} T_{\ba_1} T_{\ba_2} T_{\ba_1}^{-1} T_{\ba_2}^{-1} \ket{\psi_s} \nonumber \\
    e^{2\pi i Q^{(3)}_s} &= \arg \bra{\psi_s} C_{6}^6 \ket{\psi_s}.
\end{align}
Other possible combinations, e.g. the $C_{2,c}^2$ eigenvalue about the $c$ site, turn out to be expressible in terms of the above two invariants.
It is possible to formulate a $K$ matrix formulation for crystalline symmetry fractionalization generalizing Eq.~\eqref{eq:SF-Kmatrix}~ \cite{Manjunath2021},  but we will not use it here. \par

Mathematically, the symmetry fractionalization patterns for the CSL are classified by the cohomology group $\mathcal{H}^2(G,\Z_2)$, where $\Z_2$ is the group of anyons under fusion. 
This can be thought of as counting the number of distinct eigenvalue assignments consistent with Eq.~\eqref{eq:SF-condition}. 
When $G = \mathrm{SO}(3) \times p6$, the classification is $\Z_2 \times \Z_2 \times \Z_2 = \Z_2^3$, corresponding to the three invariants described above. 
This is reduced to  $\Z_2 \times \Z_2$  once the semion spin is fixed.\par

Refs.~\cite{Zalatel2016, Cincio2015} derive a set of symmetry fractionalization invariants analogous to the above for CSLs on the kagom\'e lattice.
They use a theoretical argument based on the `flux-fusion anomaly' test \cite{Hermele2016}, where we consider a CSL with a cylinder geometry in which two semions are created at the open ends of the cylinder by threading flux of the SO$(2)$ subgroup of SO$(3)$, and \textit{global} space group operations are performed by physically rotating the cylinder. 
Starting with a CSL on the kagom\'e lattice, it was shown that for this cylinder geometry, $Q^{(2)}_{s,cyl} = 1/2 \mod 1$ and $Q^{(3)}_{s,cyl} = 0 \mod 1$. These arguments show that there is a unique symmetry fractionalization type for the CSL on each $p6$ symmetric lattice (kagom\'e/triangular/honeycomb); in fact \cite{Song2023} generalized these arguments to $p4$ symmetric lattices as well.

Importantly, the cylinder setup differs from the one generally used to make predictions in TQFT. In particular, performing global rotations of two semions on a cylinder could introduce extra braiding factors compared to performing local rotations on an isolated semion. Therefore we should generally expect that $Q^{(j)}_{s,cyl} \neq Q^{(j)}_s \mod 1$. 
More generally, although the TQFT gives the correct \textit{classification} (i.e. the number of distinct invariants $Q^{(j)}$ and their quantization), the actual space group symmetry eigenvalues one sees in a particular experimental setup (for example $Q^{(j)}_{s,cyl}$) could be offset from the TQFT prediction by various braiding factors that appear as a result of the specific geometry of the experiment.
This issue is important because there is a close relationship between the TQFT symmetry fractionalization data and the anomaly of the CSL, discussed below.
The anomaly is what distinguishes different lattices with the same symmetry group, and will be the key CSL property that we use to test compatibility with the ordered state. 
However, the symmetry fractionalization is a more direct probe of anyon properties, and could therefore be of more experimental interest.
We might often wish to use one of these quantities to infer the other. But when we do so, we should be careful about exactly how the symmetry fractionalization is defined.  

\subsection{Anomaly}\label{sec:anomaly}

The anomaly of a global symmetry captures how it is implemented on the Hilbert space. Various notions of anomaly exist in the literature. 
Here we will define the anomaly of a $G$ symmetry to be an obstruction to having a gapped, $G$ symmetric ground state with trivial topological order. 
For a symmetry of the form $G_i \times \Gsp$ where $G_i$ is an internal symmetry, it is useful to understand anomalies as coming from Lieb-Shultz-Mattis (LSM) constraints \cite{Hastings2003LSM,Cheng2016LSM,Po2017LH,Cho2017LSM,Ye2022LSM}.
When $\Gsp$ contains only translations, we say there is an LSM constraint if the degrees of freedom in the system organize so that there is effectively a projective representation (fractional quantum number) of the internal symmetry in each unit cell. 
This notion can be refined if there are additional symmetries: for example, with $p6$ symmetry, we can separately ask if there is a projective representation at the $a,b$ and $c$ points of the unit cell. 
For $G_i = \mathrm{SO}(3)$, a nontrivial projective representation simply means half-integer spin, while a trivial projective representation means integer spin. 
It turns out that there must always be a net integer spin at each $b$ point, since there are an even number of $b$ points per unit cell. 
Hence the anomaly of $\mathrm{SO}(3) \times p6$ is given by two $\Z_2$ indices, which specify whether there are spin-$1/2$ objects at each $a$ point or at each $c$ point. 
The sum of these two indices gives whether there is an overall spin-$1/2$ representation per unit cell. Several further mathematical details relevant to the case of gapped spin liquids can be found in~\cite{Ye2022LSM}.

Why do we care about anomalies? One reason is that the anomaly is invariant under renormalization group flow, and can be used to match microscopic information to low-energy information.
Furthermore, when a system with $G$ symmetry has anomaly, it is by definition not possible to simultaneously have a ground state that is gapped, symmetric and without topological order. 
The only remaining options for the system are (i) to be gapped and symmetric but topologically ordered (as we find in the CSL), (ii) to realize a gapless ground state, or (iii) to remain gapped and without topological order but break the symmetry so that there is no longer an anomaly (as is the case in an ordered state adjacent to the CSL). 
In (i), the system can only realize the anomaly if it has a specific symmetry fractionalization pattern.
The anomaly is therefore a unifying idea that can potentially connect the universal properties of topologically ordered states to those of the ordered states that lie adjacent to it.
We will explain its precise connection to ordered states more carefully in Sec.\ref{sec: compatibility}. 

In the field-theoretic description of a topological order with symmetry, we first specify the topological order (e.g. the $K$ matrix), then the symmetry fractionalization (e.g. the charge vector). 
From these data we can in principle compute the anomaly (although this can be computationally difficult in practice). Ref.~\cite{Ye2024} previously used techniques from TQFT to compute the anomaly of each symmetry fractionalization pattern, for a large class of spin liquids. 
This analysis shows that each of the four symmetry fractionalization patterns for the CSL with $p6$ symmetry (and spin-$1/2$ semions) is associated with a different anomaly class. 
Note that this one-to-one correspondence is only a feature of the CSL topological order, in which we have just one nontrivial anyon.
For more general anyon theories, multiple symmetry fractionalization classes can map to the same anomaly.\par

For the kagom\'e CSL, the natural choice for the anomaly is that it assigns spin-$1/2$ representation of SO$(3)$ at each kagom\'e lattice site. 
(The way to formally prove this within TQFT would be to compute the symmetry fractionalization invariants for the kagom\'e CSL in the TQFT language and feed them into the formulas of \cite{Ye2024}. But as we discussed, it can be subtle to precisely define the $\Gsp$ symmetry fractionalization invariants.)

\section{Compatibility conditions between the CSL anomaly and ordered states}
\label{sec: compatibility}
 In this section we propose one general condition that an ordered state must satisfy to transition from a CSL: its topological invariant needs to be \textit{compatible} with the anomaly of the CSL, in a specific sense we now discuss. 

This condition is partly motivated by recent results on symmetry breaking in systems with anomalies \cite{Manjunath2024}. 
Consider a system with a continuous symmetry group $G$ that has an anomaly. 
Then it is possible to break the symmetry down to a subgroup $H$ which does not have anomaly.
Upon doing so, one obtains a parameterized family of gapped ground states with an order parameter living in the space $\Lambda = G/H$ \footnote{Note that if $G$ were broken spontaneously, there could be Goldstone modes, but if these are the only gapless degrees of freedom we can weakly gap them out by applying a small explicit symmetry breaking field, and by varying the order parameter value we recover the family of gapped states.}. This parameterized family has an effective action given by a nonlinear $\s$ model (NL$\s$M) for the order parameter.
The action can include various types of topological terms (often called Wess-Zumino-Witten (WZW) terms or $\theta$ terms).
The compatibility condition constrains the coefficients of these terms based on the original $G$ anomaly. 

There is a large prior body of work studying topological terms in NL$\s$Ms without referring to a $G$ anomaly; see for instance Ref.~\cite{Abanov1999-NLSM}. Ref.~\cite{Manjunath2024} gave a mathematical procedure to determine the compatibility relation (i.e. the constraint between the $G$ anomaly and the topological terms in the nonlinear $\s$ model for the family) by relating the classification of anomalies to that of parameterized families, without explicitly writing out the field theories.
This procedure is summarized in Appendix~\ref{app:families}.
Although abstract, this analysis has the advantage that physical information (such as the properties of order parameter defects) can be determined even when a field-theoretic formulation could be complicated to work with.
To our knowledge, the mathematical details in Appendix~\ref{app:families} represent the first concrete connection drawn between parameterized families and magnetically ordered spin systems. \par

\subsection{The basic compatibility condition for kagom\'e CSL}
The kagom\'e CSL anomaly implies that there is effectively a spin-$1/2$ degree of freedom localized at each kagom\'e site, while there is no spin-$1/2$ degree of freedom at the triangular sites. This means that upon breaking symmetry, the ordered state must also be organized so that \\
\begin{itemize}
\item Each kagom\'e site has a magnetically ordered degree of freedom which transforms as a half-integer spin object under SO$(3)$; \\
\item The degrees of freedom on the triangular sites, if any, transform as integer spin objects. \\
\end{itemize}
Additionally, since there are effectively an even number of spins on the honeycomb lattice per unit cell, we necessarily have a net integer spin and a trivial anomaly contribution from those sites. These conclusions can be recovered from the mathematical analysis in Appendix~\ref{app:families}.

We note a couple of points. First, this is not a hugely restrictive condition as RMOs on the kagom\'e lattice would be compatible with the kagom\'e lattice anomaly, as long as the ordered variables all transform as spin-$1/2$ objects. 
Note, however, that several classical RMOs discussed in \cite{Messio2011}, such as both types of cuboctahedral order, cuboc-1 and cuboc-2, cease to be RMOs when considering quantum spins with broken TRS, as discussed in Section \ref{sec: RMO}.
This can easily be seen by considering the scalar spin chirality of the cuboc orders on various triangular plaquettes on the kagom\'e lattice which flip under lattice operations such as translation or rotation. 
Second, these conditions are necessary to allow a continuous transition from the CSL into an ordered state but they are not sufficient.
One can imagine adding perturbations to the Hamiltonian that force the system to order through a first-order transition. 
The following examples show how the compatibility condition does give some nontrivial information.

\subsection{Example: Ferromagnet to chiral spin liquid}

Let us return to the example of the ferromagnet. Recall that we characterized the ferromagnet in terms of three integers $(C_a, C_b, C_c)$ which describe the Berry-Chern number per unit cell of magnetized spins at the $a,b,c$ sites respectively. 
Now in the zero-dimensional case of a single spin $s$ with a single Berry-Chern number $C$, one can show that $C$ must be odd whenever $s$ is a half-integer, while $C$ must be even whenever $s$ is an integer. The reason is that $C = 2m$ where $m$ is the $S_z$ eigenvalue of each magnetized spin state, and $m = s \mod 1$. The two compatibility conditions therefore read as
\begin{itemize}
    \item The CSL anomaly assigns spin-$1/2$ to each kagom\'e site $c$. Therefore $C_c$ must be odd. 
    \item The CSL anomaly assigns integer spin to each triangular site $a$. Therefore $C_a$ must be even.
\end{itemize}
Furthermore, the CSL anomaly does not constrain the spin at the honeycomb sites $b$; therefore $C_b$ can be arbitrary.

\subsection{Example: Noncoplanar order to chiral spin liquid}

We now focus on the `melting' of noncoplanar crystals through a continuous transition to a chiral spin liquid. 
The compatibility condition between these two states requires that the projective degrees of freedom of the CSL, the semions, form the same lattice as the spin-$1/2$ degrees of freedom forming the ordered state. This condition can easily be satisfied for the noncoplanar orders described in Table \ref{table: Noncoplanar RMOs}.
Since this is a rather generous compatibility condition, it allows us consider possible continuous transitions between these states. 
Another intuitive appeal of focusing on such a transition is the fact that both states have broken chiral symmetry.
Furthermore, we focus on the octahedral order to CSL on the kagom\'e as it is one possible explicit example \cite{Bose2023}, although we comment where the analysis is more general.

\par
It is interesting to note that in the symmetry group of the octahedral order, translations anticommute while the $\tilde{C}_6$ rotation satisfies $\tilde{C}_6^6=+1, \tilde{C}_{2,c}^2 = -1$. We know that in the CSL, the translation symmetry is fractionalized, meaning that translations anticommute on a state with a semion.
Furthermore, in order to be consistent with the kagom\'e lattice anomaly, the $C_6$ rotation symmetry should not be fractionalized (as per the definitions of Ref.~\cite{Ye2024}) \footnote{Although the results of Refs.~\cite{Zalatel2016,Cincio2015,Song2023} would seem to imply that the $C_6$ symmetry \textit{is} fractionalized, these works are likely using a  different definition compared to \cite{Ye2024}, as mentioned in Sec.\ref{sec:CSL-review}.}.
Does this naive similarity between the symmetries of the octahedral state and the symmetry fractionalization in the kagom\'e CSL imply that there is a more likely case for a continuous transition? 
We believe the answer is no, because the tetrahedral orders on the triangular/honeycomb lattices have precisely the same group relations, but we expect the symmetry fractionalization to be different in those cases \cite{Song2023}. 

 In this section we have argued that a compatibility relation should exist without explicitly writing out any field theory. 
 However, we expect it should be possible to recover the above conclusions within field theory, as follows.
 We start from an effective response theory only involving background gauge fields for the $G$ symmetry, which encodes the anomaly of the CSL.
 Then we break the symmetry so that the background $G$ gauge field is replaced by an order parameter field valued in $\Lambda$ and a gauge field for the symmetry $H_0$. 
 The presence of a topological term corresponding to the anomaly of the CSL places a constraint on the allowed topological terms for the order parameter, giving us the desired compatibility relation. We do not use this approach for the octahedral spin crystal since it is not obvious how to formulate a topological term for an $\mathbb{RP}^3$ order parameter; we prefer to directly work with the mathematical classification instead, as we discuss in Appendix~\ref{app:families}.

\section{Field Theory Description of the CSL to noncoplanar Transition}

\label{sec: field theory}

In previous sections we have addressed our general objective by establishing compatibility of the CSL with the regular anti-ferromagnetic noncoplanar orders listed in Table \ref{table: Noncoplanar RMOs} (octahedral order on the kagom\'e, or tetrahedral order on the triangular and the honeycomb lattice). In the following section we wish to tackle our specific objective of constructing a quantum field theory which can describe the transition between said states. 
The symmetry group we are considering is \(G=p6\times \text{SO}(3)\) which does not include time-reversal or reflection symmetries, as these are assumed to be broken explicitly in the microscopic system. 
This is the case, for example, when including an explicit chiral interaction $H = - J_{\c} \sum_\triangle (\bS_i \cdot \bS_j \times \bS_k)$ in the microscopic spin model. Furthermore, for convenience, the basis for lattice space group symmetries in this section will be chosen to be \(\{T_{\ba_1}\,,T_{\ba_2}\,,C_{2,a}, C_{3,b}\}\) as opposed to \(\{T_{\ba_1}\,,T_{\ba_2}\,,C_{6,a}\}\). We derive the action from a general symmetry analysis of the continuum degrees of freedom and justify the existence of the two phases in that manner, rather than through a specific microscopic spin Hamiltonian using the coherent state spin path integral. 
This is a similar approach to previous derivations of the field theoretic description for a transition between a coplanar order and a \(\Z_2\) spin liquid \cite{Sachdev1994, Sachdev2009}. \par 

As mentioned previously, the CSL can be expressed in terms of Gutzwiller projected spin-$1/2$ fermionic partons filling Chern bands with a total Chern number  $C_\upa + C_\dna =2$ \cite{Ran2007}. 
The chiral spin liquid then has a descendant low energy theory of gapped \(N_f=4\) Dirac fermionic partons coupled to an emergent U$(1)$ gauge theory 
\begin{equation}
    \label{dirac theory}
    \mathcal{L}=\sum_{i=1}^4 \bar{\psi}_i (i \slashed{D}-m) \psi_i+\frac{1}{4} f_{\mu \nu}^2\,,
\end{equation}
where $\slashed{D} = \gamma^\mu\left(\partial_\mu-i a_\mu\right)$ and $f_{\mu \nu} = \epsilon^{\mu \nu \lambda} \partial_\nu a_\lambda$. 
Due to the chiral mass $m \overline{\psi}_i \psi_i$, the gapped fermions can be integrated out completely.
One is left with a pure $\mathrm{U}(1)_2$ Chern-Simons gauge theory as a result
\begin{equation}
 \mathcal{L}_{\mathrm{CSL}}  =  \frac{2}{4\pi} a \wedge \text{d} a\,,
\end{equation}
which hosts two topological excitations : a trivial \(e\) and a semion \(s\), as well as a chiral edge mode.\par
Transitions out of the CSL can be parameterized with symmetry breaking masses, such as done for a CSL to Ne\'el transition \cite{Samajdar2019}.
In this language, the noncoplanar ordered state corresponds to a matrix mass in flavor space \(n_i^a=\overline{\psi} \cdot \tau^i \otimes \s^a \cdot \psi\) \cite{Song2019, Song2020}. 
Therefore, the transition from the CSL to the ordered state could correspond to tuning two masses: the chiral mass and the matrix mass such that the critical point corresponds to a Dirac spin liquid with a low energy theory described by a QED\(_3\) action \cite{Hermele_2004_Stability, Hermele2005_ASL}. 
However, once symmetries are broken by the order parameter, monopole proliferation can occur leading to complicated bond ordering coexisting with magnetic ordering \cite{Song2019, Song2020}. \par

To avoid such technical difficulties, we instead work with bosonic theories. Simple collinear orders like the N\'eel order have been studied in the language of \(O(3)\) nonlinear sigma-model (NL\(\s\)M) written in terms of a slowly varying N\'eel vector \(\hat{n}(x)\) in the long-wavelength limit \cite{dombre1988, Sachdev2011, Fradkin_2013}. In the disordered phase, the local Berry phase term in the spin path integral becomes relevant and can provide topological properties of the phase \cite{dombre1988, Fradkin1988, Haldane1988}.
This topology becomes more manifest by fractionalizing the N\'eel vector into bosonic spinons coupled to a \(\mathrm{U}(1)\) gauge field which leads to the \(\mathbb{CP}^1\) formulation \cite{Haldane1988, Fradkin1988}. As we shall see in the subsequent sections, the generalization of the \(\mathbb{CP}^1\) construction to multiple flavors, as well as to \(\mathrm{SU}(2)\) matrix equivalents, provides us the natural connection between fractionalization in the CSL and the noncoplanar ordered state.

\subsection{ Nonlinear sigma-model for the ordered state}

The octahedral and the tetrahedral orders are characterized by three orthogonal unit vectors \(\hat{n}_i^a\) at the three \(\bM\) points in the Brillouin zone. We refer to orders defined by three characteristic wave-vectors as triple-Q orders. For the triangular and the honeycomb lattice the orthogonal vectors live on the same lattice site while for the kagom\'e they each occupy different sublattices as defined in Table \ref{table: Noncoplanar RMOs expectations}. The ordering pattern can also be thought of as three N\'eel orders on deformed square lattices, forced to be mutually orthogonal to each other.\par

\begin{table*}[!ht]
\centering
\begin{tabular}{ |c|c|c|c| } 
     \hline
     Order & Lattice & Sublattice : \(i\) & \(N_i^a(\br)\)\\
     \hline
     Tetrahedral & Triangular & \(1\) & \(\left(1/\sqrt{3}\right)\left(\hat{n}_1^a(\br)\cos(\bM_1\cdot \br)+\hat{n}_2^a(\br)\cos(\bM_2\cdot \br)+\hat{n}_3^a(\br)\cos(\bM_3\cdot \br)\right)\)\\
     \hline 
     \multirow{2}{*}{Tetrahedral} & \multirow{2}{*}{Honeycomb} & \(1\) &\(\left(1/\sqrt{3}\right)\left(\hat{n}_1^a(\br)\cos(\bM_1\cdot \br)+\hat{n}_2^a(\br)\cos(\bM_2\cdot \br)+\hat{n}_3^a(\br)\cos(\bM_3\cdot \br)\right)\)\\ \cline{3-4}
     & & \(2\) & \(\left(1/\sqrt{3}\right)\left(-\hat{n}_1^a(\br)\cos(\bM_1\cdot \br)-\hat{n}_2^a(\br)\cos(\bM_2\cdot \br)+\hat{n}_3^a(\br)\cos(\bM_3\cdot \br)\right)\)\\
     \hline 
     \multirow{3}{*}{Octahedral} & \multirow{3}{*}{kagom\'e} & \(1\) & \(\hat{n}_1^a(\br) \cos(\bM_1\cdot \br)\)\\ \cline{3-4}
     & & \(2\) & \(\hat{n}_2^a(\br) \cos(\bM_2\cdot \br)\)\\ \cline{3-4}
     & & \(3\) & \(\hat{n}_3^a(\br) \cos(\bM_3\cdot \br)\)\\ \hline
\end{tabular}
\caption{The spin expectation value \(\expval{S_i^a(\br)} = (1/2)N_i^a(\br)\) in the coherent spin path integral expressed in terms of three unit vectors \(\bn_i\) at the three \(\bM\) points for the three noncoplanar regular antiferromagnetic orders with \(p6\times \mathrm{SO}(3)\) symmetry.
All these three orders can be represented by an orthonormal trial of unit vectors \(\hat{n}_i^a\), or equivalently by a reference frame in spin-space.}
\label{table: Noncoplanar RMOs expectations}
\end{table*}

The unit vectors defining the order transform as follows under spin and spatial transformations
\begin{itemize}
    \item Spin-Rotation : Under a spin-rotation \(\mathcal{R}\),  \(\hat{n}_i\) must transform as a SO$(3)$ vector, \(\hat{n}_i^a \xrightarrow{\mathcal{R}} \mr_{ab}\cdot\hat{n}_i^b\) where \(\mr\in \mathrm{SO}(3)\). 

    \item Time-Reversal : Under time-reversal \(\mathcal{T}\), all spins must flip and hence \(\hat{n}_i \xrightarrow{\mathcal{T}} -\hat{n}_i\). 
    
    \item Lattice Translation : Under lattice translations \(T_{\ba_{i}}\,,i\in\{1, 2\}\), the spin orientations should transform as \(\hat{n}_i(\br)\xrightarrow{T_{\ba_j}}\hat{n}_i(\br+\ba_j) = \hat{n}_i(\br)\cos(\mathbf{M}_i\cdot\ba_j)\). 

    \item Lattice Rotation : Under a \(C_3^b\) lattice rotation, \(\hat{n}_i\) are interchanged cyclically. Meanwhile under a \(C_2^a\), they are invariant.
\end{itemize}

The orthogonality constraint of \(\hat{n}_i\cdot\hat{n}_j = \d_{ij}\) can be imposed exactly using a complex vector boson parametrization which fractionalizes the unit vectors.
However, this parametrization reduces the gauge degree of freedom from \(\mathrm{U}(1)^3\) down to \(\Z_2\) and was used to describe transitions into a \(\Z_2\) spin liquid \cite{Angelucci1991, Sachdev1994}. Since we want to describe the transition to a chiral spin liquid which is described by U$(1)$ Chern-Simons theory, we require a different approach. A possible route is to soften the orthogonality constraint and formulate the low energy theory on the ordered side as three \(O(3)\) nonlinear sigma-models (NL$\s$M) coupled to each other through the chiral term as 
\begin{equation}
    \label{coupled sigma models}
    \ml_{n} = \frac{1}{2g}\sum_{i}\left(\partial_{\m}\hat{n}_i\right)^2 - \c \left(\hat{n}_1 \cdot \hat{n}_2 \times \hat{n}_3\right)\,,
\end{equation}
which preserves all of the aforementioned symmetries except time-reversal. Each stiffness parameter \(g_i \equiv g \) is related by \(C_3^b\) symmetry and  must be equal to each other. Note that there are no terms which couple the NL$\s$Ms to each other at \(\mo(n^2)\). Terms of the form \(\hat{n}_i\cdot \hat{n}_j\) break the implementation of lattice translation symmetry for the triple-Q noncoplanar order. Such terms would be rapidly oscillating at the lattice scale and  do not appear in the long-wavelength limit in which the continuum field theory is written. Furthermore, since we are only writing down the NL\(\s\)M on the ordered side, the Berry phase term does not play any role in the dynamics and can be safely dropped in the action in Eq.\eqref{coupled sigma models}. These are the terms that would specify the parameterized family on the ordered side, as we noted in Sec.~\ref{sec: RMO}.

In the ordered phase (\(g\ll 1\)), the \(\hat{n}_i\) fields condense and gain a vacuum expectation value \(\expval{\hat{n}_i}\neq 0\). Without the chiral interaction the three condensates are arbitrary and independent from each other. The chiral interaction forces them to be in a configuration which maximizes the chirality \(\expval{\hat{n}_1}\cdot \expval{\hat{n}_2} \times \expval{\hat{n}_3} = 1\), which imposes orthogonality as \(\expval{\hat{n}_i}\cdot \expval{\hat{n}_j}=\d_{ij}\). Further proof that this indeed describes the desired noncoplanar ordered phase can be obtained by analyzing the low energy transverse excitations around the condensate \(\hat{n}_i = \expval{\hat{n}_i}\sqrt{1-\Tilde{n}_i^2} + \Tilde{n}_i\) with \(\expval{\hat{n}_i}\cdot\Tilde{n}_i=0\). 
Substituting this back into Eq.\eqref{coupled sigma models}, the Lagrangian to leading order reduces to (up to the constant energy of the condensate)
\begin{equation}
    \label{Lagrangian leading order}
    \ml = \frac{1}{2g}\sum_{i}(\partial_{\m}\tilde{n}_i)^2+\c\left( \frac{1}{2}\sum_{i}\tilde{n}_i^2 + \tilde{n}_1^y\tilde{n}_2^x+\tilde{n}_2^z \tilde{n}_3^y+\tilde{n}_3^x\tilde{n}_1^z\right)\,.
\end{equation}
The chiral interaction can be shown to gap out 3 out of the \(3\times2=6\) zero modes expected from three \(O(3)\) NL$\s$Ms. This matches the spin-wave result expected in the octahedral or the tetrahedral state where SO$(3)$ is broken completely and the counting is consistent with previously known results \cite{Watanabe2020}. \par 

In the next section we connect this theory of the ordered state to that of a CSL through the \(\mathbb{CP}^1\) formulation which fractionalizes the \(O(3)\) unit vectors into bosonic spinors coupled to an \emph{emergent} gauge field. These emergent gauge degrees of freedom produce a topological term. While the term does not affect the ordered side \cite{Larkin1988, Haldane1988, dombre1988, wen1988, Fradkin1988}, we find that the combined cluster of \(\mathbb{CP}^1\) theories conspires to provide a Chern-Simons theory on the disordered side describing a $\mathrm{U}(1)_2$ CSL. Although the formal connection to the fermionic theory coupled with \(\mathrm{U}(1)\) gauge fields referred in Eq.\eqref{dirac theory} is unclear, we will find the effective low energy theory to have the same symmetries and topological content in the \(\mathbb{CP}^1\) language. \par

\subsection{Multi-component CS Theory for U$(1)_2$ CSL}

In this section, we fractionalize the theory in Eq.\eqref{coupled sigma models} using bosonic spinons coupled to emergent U$(1)$ gauge fields. The \(O(3)\) unit vector fields can be parametrized using bosonic spinors as
\begin{equation}
    \label{fractionalization}
    \hat{n}_i^a(x) = z^{\dagger}_{i, \a}(x)\s^{a}_{\a\b}z_{i, \b}(x)\,,\:\:\:z^{\dagger}_{i, \a}z_{i, \a}=1\,,
\end{equation}
which has a gauge redundancy : the transformation \(z_{i, \a}(x) \rightarrow e^{i\phi^i(x)}z_{i, \a}(x)\) leaves the unit vectors invariant. One can write a Lagrangian in terms of the bosons \(z_i\) minimally coupled with the gauge fields as
\begin{equation}
    \label{CP1 action}
    \ml_{z} = \frac{1}{g}\sum_{i}|D_{\m}^{i}z_{i}|^2 - \c \left(\hat{n}_1 \cdot \hat{n}_2 \times \hat{n}_3\right)\,,\:\:\:D_{\m}^{i}=\partial_{\m}-i a^{i}_{\m}\,,
\end{equation}

This $\mathbb{CP}^1$ model is exactly equivalent to the coupled NL$\s$M in Eq.\eqref{coupled sigma models} which we expect in the symmetry broken phase \cite{Fradkin_2013}. 
This provides a good starting point to relate to the CSL phase coming from the ordered side.
Next, we introduce topology into the action through a  Chern-Simons term in the $\mathbb{CP}^1$ gauge fields \cite{Wu1984, Voruganti1988}. 
Generically, since we have three ``flavors" of U$(1)$ gauge fields $a_{\m}^{i}$, the Chern-Simons term comprises of a \(3\times3\) \(K\)-matrix. The full Lagrangian has the form
\begin{equation}
    \label{eq:field-theory}
    \begin{gathered}
            \ml = \frac{1}{g}\sum_{i}\left|D_{\m}^{i} z_i\right|^2 - \chi\left( \hat{n}_1\cdot \hat{n}_2\times\hat{n}_3\right) \\
            + \e^{\a\b\g} \sum_{i,j} \frac{K_{ij}}{4\pi} a^i_{\a}\partial_{\b}a^j_{\g} + \cdots\,
    \end{gathered}
\end{equation}
where the \(K\)-matrix must be symmetric, have integer entries, and have even integers for the diagonal entries since we are dealing with bosonic spins. 
Note that this field theory does not precisely encode the underlying lattice- it only incorporates the $p6$ symmetry.
The distinction between different lattices will appear through additional terms in $\cdots$ that can be made explicit if one includes background gauge fields for the crystalline symmetry. 
On the ordered side, such terms would distinguish the octahedral order on the kagom\'e lattice and the tetrahedral order on the triangular lattice. On the disordered side, the same terms would specify the lattice anomaly of the CSL, in accordance with the compatibility relation of Sec.~\ref{sec: compatibility}. Although we will not require them here, for completeness in Appendix~\ref{app:anomalyterms} we briefly review how to formulate the CSL anomaly terms within field theory.

\par

Using the transformation properties of the \(O(3)\) vectors \(\hat{n}_i^a\) derived in the previous section, we can derive the spinon transformations as follows:
\begin{itemize}
    \item Spin-Rotation : Under a spin-rotation \(\mathcal{R}\), the spinons transform as spin-$1/2$ particles \(z_i \xrightarrow{\mathcal{R}} \mathcal{U}(\mr)\cdot z_i\) where \(\mathcal{U}(\mr)\in \mathrm{SU}(2)\). 

    \item Time-Reversal : Under time-reversal \(\mathcal{T}\), the bosonic spinon transforms as \(z_i \xrightarrow{\mathcal{T}} \mathcal{K}\cdot i\s^y z_i\) where \(\mathcal{K}\) is the anti-linear complex conjugation part of time-reversal.
    
    \item Lattice Translation : Under lattice translations \(T_{\ba_{i}}\,,i\in\{1, 2, 3\}\)
    \begin{equation}
        \label{translation kagom\'e}
        \begin{gathered}
            z_1 \xrightarrow{T_{\ba_1}} i\s^y z_1^{*}\,,\:\:\: z_2 \xrightarrow{T_{\ba_1}} z_2\,,\:\:\: z_3 \xrightarrow{T_{\ba_1}} i\s^y z_3^{*}\,,\\
            z_1 \xrightarrow{T_{\ba_2}} z_1\,,\:\:\: z_2 \xrightarrow{T_{\ba_2}} i\s^y z_2^{*}\,,\:\:\: z_3 \xrightarrow{T_{\ba_2}} i\s^y z_3^{*}\,.
        \end{gathered}
    \end{equation}

    \item Lattice Rotation : Under a \(C_3^b\) lattice rotation the spinon flavors are interchanged cyclically, while they are invariant under \(C_2^a\).
\end{itemize}

Now when we demand that the non-topological part of Eq.\eqref{eq:field-theory} stays invariant under the full \(p6\times \mathrm{SO}(3)\) symmetry, we find the symmetry action on the gauge fields \(a_{\m}^{i}\) which constrains the allowed \(K\)-matrices. For instance, \(C_3^b\) symmetry forces \(K= a \mathds{1}+b M\) where \(a, b\) are arbitrary integers of the same parity, and \(M\) is the \(3\times3\) matrix with all entries being 1.
Furthermore, as discussed in Sec.\ref{sec:anomaly}, we already know the topological content of the abelian KL-CSL which hosts two topological excitations : the trivial \(e\) and the semion \(s\), along with one chiral edge mode. Hence by the nature of transition we want to describe, we require that \(|\det(K)|=2\), and \(|\mathrm{sgn}(K)|=1\). With the help of symmetry constraints, one can formulate the \(K\)-matrix and the semion charge vector \(s\) with the required properties as
\begin{equation}
    \label{CSL k matrix}
    K = -\begin{pmatrix}
        0 & 1 & 1\\
        1 & 0 & 1\\
        1 & 1 & 0
    \end{pmatrix}\,,\:\:\: s = \begin{pmatrix}
        1 \\
        1 \\
        1
    \end{pmatrix}\,,\:\:\:s^{T}\cdot K^{-1} \cdot s = \frac{1}{2}\,.
\end{equation}

Some comments about this form of the \(K\)-matrix are in order. The \(K\)-matrix preserves lattice rotation symmetry \(C_3^b\) and \(C_2^a\) explicitly but naively breaks translation symmetry since only two out of three of the gauge fields flip under a given translation direction.  This can be resolved by considering a theory of the form Eq.\eqref{eq:field-theory} with a given \(K\)-matrix as dual to another with \(K' = G^{T}\cdot K \cdot G\) where \(G\in \mathrm{GL}(3, \Z)\) since they have the same topological content \cite{ Lu2012, Lu2016, Benini2018, Shyta2022}. This is indeed possible for the action of translation symmetry where the \(\mathrm{GL}(3, \Z)\) transformations can be chosen as \(G_{\ba_1} = \mathrm{diag}(-1, 1, -1)\) and \(G_{\ba_2}=\mathrm{diag}(1, -1, -1)\). Moreover, this symmetry implementation which relies on dualities turns out to be unnecessary in the matrix parton formulation that we discuss in Sec. \ref{sec: matrix parton}. \par 

The \(K\)-matrix we derive can also be shown to be equivalent up to a \(\mathrm{GL}(3, \Z)\) transformation to another matrix \(K'\) with  \(K'=2 \oplus (-\sigma^x)\). 
Hence, the system can be interpreted as a CSL coexisting with a background state with trivial anyon content (the sub-space spanned by \(\s^x\)). 
The topological sector is the uniform linear combination of all the three gauge fields as evident from the semion charge in Eq.\eqref{CSL k matrix}, while other orthogonal combinations end up being trivial. \par

Now we have all the pieces of the puzzle.
We constructed an action for three flavors of spinons coupled to three U$(1)$ gauge fields with a Chern-Simons term which respects all the symmetries in \(p6\times \mathrm{SO}(3)\) up to topological dualities enacted through \(\mathrm{GL}(3,\Z)\) transformations.
The theory can be tuned to criticality by tuning the value of \(g\). When \(g<g_{c}\), all three spinons condense simultaneously, which leads to symmetry breaking and Higgsing of the gauge fields. As a result, the Chern-Simons term gets gapped which allows us to integrate out the photons and recover three \(O(3)\) NL$\s$Ms coupled via a chiral interaction ala Eq.\eqref{coupled sigma models}. This is the magnetically ordered phase which breaks SO$(3)$. Meanwhile, when \(g>g_c\), the spinons are gapped and hence can be integrated out leaving us with just the Chern-Simons theory with the same topological properties as the CSL. 
At criticality, the chiral interaction is \(\mo(z^6)\) is RG-marginal by power counting as the theory lives in $2+1$D. The critical theory therefore reduces to interacting massless bosonic spinons coupled to a Chern-Simons gauge theory. \(\mo(z^4)\) symmetry-allowed perturbations involving derivatives such as \(\sum_{i=1}^{3}|z^{\dagger}_i\cdot \partial_{\m}z_i|^2\) will be irrelevant at tree-level. It is important to note that even if a term like this becomes relevant in the strongly coupled theory, it \emph{will not} change the symmetry broken order and only renormalize the spin-wave modes. Furthermore, one can show that no \(\mo(z^4)\) term can be generated through gauge fluctuations which are independent of momentum and are incapable of changing the nature of the ordering. These terms can be neglected at criticality  as well.\par

\subsection{Matrix parton theory} \label{sec: matrix parton}
In this section, we propose a more universal formalism to treat transitions between noncoplanar orders and topologically ordered states in terms of matrix partons. While the field theory discussed in the previous section has a physically appealing interpretation of three \(O(3)\) NL\(\s\)M coupled via a chiral interaction, it has the drawback of requiring multiple flavors of gauge fields, as well as the requirement to appeal to dualities while implementing lattice symmetries. Furthermore, the constraint that the three unit vectors \(\hat{n}_i\) are orthogonal to each other is only imposed as a soft constraint through the addition of a RG-irrelevant chiral interaction. In this section instead, we shall formulate a field theory which resolves all these aforementioned issues. The principal motive is to extend the idea of fractionalizing a unit vector order parameter of a simple collinear order \(\hat{n}\), to noncoplanar order parameters defined by a orthonormal triad of vectors \(\hat{n}_i\,,i\in\{1, 2, 3\}\) with \(\hat{n}_i\cdot \hat{n}_j = \d_{ij}\). Such a orthonormal triad can be re-parametrized using a rotation matrix \(R\in \mathrm{SO}(3)\) acting on a frame of reference for the noncoplanar order. This is equivalent to the \(\mathbb{RP}^3\) order parameter discussed in Sec. \ref{sec: RMO} for family classification. Subsequently, when this rotation matrix is assumed to have long-wavelength fluctuations, a field theory can be written down.\par

Now, a unit vector transforming under $\mathrm{SO}(3)$ is fractionalized in terms of $\mathrm{SU}(2)$ spinors as \(\hat{n}^a = z^{\dagger}\cdot \s^a \cdot z\), and the low energy theory is written in terms of the spinons \(z\) in the usual \(\mathbb{CP}^1\) theory.
Here, we instead want to fractionalize a SO$(3)$ matrix and the natural extension is to use \(\mathrm{SU}(2)\) matrices:
\begin{equation}
    \label{matrix partons}
    \hat{n}_i^a \equiv R_{ab} e_i^b = u_i^{\dagger}\cdot U^{\dagger}\cdot \s^a \cdot U \cdot u_i\,,\:\:\:U^{\dagger}\cdot U=\mathds{1}\,,
\end{equation}
where \(u_i=\ket{e_i, +}\) are three normalized reference spinors along three reference orthonormal $\mathrm{SO}(3)$ vectors \(e_i^a\) with \(\hat{e}_i\cdot \hat{e}_j = \d_{ij}\) and \(\hat{e}_1\cdot (\hat{e}_2 \times \hat{e}_3) = 1\). We can choose a convention where the reference vectors are fixed to be \(\hat{x}, \hat{y},\) and \(\hat{z}\) on the Bloch sphere such that \(e_i^a = u_i^{\dagger}\cdot \s^a \cdot u_i = \d_i^a\) for simplicity, but we report the general results.
Now, Eq.\eqref{matrix partons} clearly has a U$(1)$ gauge redundancy since \(U \rightarrow e^{i\phi}U\) does not affect the physical order parameter \(R\). Individual unit vectors might be unaffected by a larger gauge group \textit{e.g.} \(\hat{n}_1\) is also invariant under \(U\rightarrow U\cdot \exp\left(i \phi \hat{e}_1\cdot \boldsymbol{\s}\right)\). However, if we require \emph{all three} unit vectors to stay invariant, their orthogonality forces the common gauge group to be reduced to just \(\mathrm{U}(1)\). Hence, we  treat the parton as living in \(U \in \mathrm{U}(2)\) with a U$(1)$ gauge degree of freedom.\par
As before, we can work out all the actions of \(G=p6\times \mathrm{SO}(3)\) on these partons :
\begin{itemize}
    \item Spin-Rotation : Under a spin-rotation \(\mathcal{R}\in \text{SO}(3)\), Each \(\hat{n}_i\) must transform as a SO$(3)$ vector, \(\hat{n}_i^a \xrightarrow{\mathcal{R}} \mathcal{R}_{ab}\cdot\hat{n}_i^b\). In terms of the matrix fields \(U\), this implies \(U\xrightarrow{\mathcal{R}} \mathcal{U}(\mathcal{R})\cdot U\) where \(\mathcal{U}(\mathcal{R})\in \text{SU}(2)\). 

    \item Time-Reversal : Under time-reversal \(\mathcal{T}\), All spins must flip and hence \(\hat{n}_i \xrightarrow{\mathcal{T}} -\hat{n}_i\). For the \(\text{SU}(2)\) fields this is implemented as \(U \xrightarrow{\mathcal{T}} \mathcal{K}(i\s^y)\cdot U\) where \(\mathcal{K}\) is the anti-unitary complex conjugation part of time-reversal.
    
    \item Lattice Translation : Lattice translations are accomplished by \(U \xrightarrow{T_{\ba_i}}U\cdot \mathcal{U}_{T_{\ba_i}}\) with
    \begin{equation}
        \label{translation kagom\'e U}
        \mathcal{U}_{T_{\ba_1}} = \exp\left(i\frac{\pi}{2}\hat{e}_2\cdot\boldsymbol{\s}\right)\,,\:\:\:\mathcal{U}_{T_{\ba_2}} = \exp\left(i\frac{\pi}{2}\hat{e}_1\cdot\boldsymbol{\s}\right)\,.
    \end{equation}

    \item Lattice Rotation : A \(C_{6}^a\) lattice rotation is implemented by \(U \xrightarrow{C_6^a}U\cdot \mathcal{U}_{C_6^a}\) with
    \begin{equation}
        \label{rotation kagom\'e U}
        \mathcal{U}_{C_6^a} = \exp\left(i\frac{2\pi}{6}\hat{n}_0\cdot \boldsymbol{\s}\right)\,,\:\:\:\hat{n}_0 = \frac{1}{\sqrt{3}}(-\hat{e}_1+\hat{e}_2+\hat{e}_3)\,.
    \end{equation}
\end{itemize}
These symmetries have been specified for the regular antiferromagnetic orders mentioned in Table \ref{table: Noncoplanar RMOs} but in general can be replaced appropriately for \emph{any} ordering. In fact the symmetry implementation can be directly read off from the residual symmetry of the ordered state listed in Table \ref{table: Noncoplanar RMOs}, Table \ref{table: coplanar and collinear RMOs}, and Table \ref{table: ferro collinear RMOs}. \par

One can now construct a field theory based on these symmetries as
\begin{equation}
    \label{matrix field theory}
    \ml_U = \frac{1}{g}\Tr\left(D_{\m} U^{\dagger}\cdot D_{\m} U\right) + \frac{K}{4\pi}\e^{\m\n\l}a_{\m}\partial_{\n}a_{\l}+...\,,
\end{equation}
where \(...\) refers to higher order interactions which are allowed to break TRS but are irrelevant at criticality. The Lagrangian in Eq.\eqref{matrix field theory} is symmetric under the whole \(p6\times \mathrm{SO}(3)\) symmetry group explicitly without the need for duality. The free kinetic term \(\Tr\left(\partial_{\m}U^{\dagger}\cdot\partial_{\m}U\right)\) in the Lagrangian is sometimes referred to as the principal chiral model \cite{Batista2018}. \par 

The full continuum theory has an enlarged symmetry group of \(\mathrm{U}(2)_L\times \mathrm{U}(2)_R\) which may be recast as \(\mathrm{U}(1)_{g}\times \mathrm{SU}(2)_L\times \mathrm{SU}(2)_R\) symmetry where \(L\) and \(R\) refer to left and right multiplication. The physical spin-rotation SO$(3)$ symmetry gets mapped to \(\mathrm{SU}(2)_L\), the lattice symmetries form a non-abelian subgroup \(p6\subset \mathrm{SU}(2)_R\), and the U$(1)$ symmetry is being gauged. The gauge field can generically have a Chern-Simons term since time-reversal is assumed to be broken explicitly. 
The only constraint on the coefficient \(K\) is that it should be even, \(K\in 2\mathds{Z}\). Different values of \(K\) can describe transitions into different \(\mathrm{U}(1)_{K}\) topological orders. \par

Lastly, let us try to relate the matrix parton \(U\in \mathrm{U}(2)\) to the usual SU$(2)$ spinors. Just by counting the degrees of freedom, we know \(U\) has three real fields to define a SU$(2)$ rotation plus one gauge degree of freedom. Now a normalized SU$(2)$ spinor also has \emph{only} three real parameters. This is exactly the reason why in previous works the residual gauge symmetry for the spinor was reduced to  \(\mathds{Z}_2\) \cite{Sachdev1994, Sachdev2009}. To see an effective U$(1)$ gauge, one is forced to introduce a complex scalar field \(\Phi\) which is purely gauge.
A convenient way is as follows
\begin{equation}
    \label{U2 parametrization}
    U = \Phi\begin{pmatrix}
        z_{\ua} & -z_{\da}^{*}\\
        z_{\da} & z_{\ua}^{*}
    \end{pmatrix}\,,\:\:\:z^{\dagger}z=1\,,\:\:\: |\Phi|^2=1\,.
\end{equation}
This is composed of a two-component spinor \(z\) and a complex scalar \(\Phi\), both of which are norm-1. The gauge transformation \(U \rightarrow e^{i\phi} U\) is realized by \(\Phi\rightarrow e^{i\phi}\Phi\).  \(z\) is a charge-zero SU$(2)$ spinor while \(\Phi\) is a charge-one scalar. The different spinors fractionalizing the vectors \(\hat{n}_i\) are related to the aforementioned fields as
\begin{equation}
    \label{spinor relation}
    z_1 = \frac{\Phi}{\sqrt{2}}\left(z+i\s^y z^{*}\right)\,,\:\:\:z_2 = \frac{\Phi}{\sqrt{2}}\left(z+\s^y z^{*}\right)\,,\:\:\:z_3=\Phi z\,.
\end{equation}
In fact, the spinor \(z\) and charged-scalar \(\Phi\) can even be related to the bosonic partons \(S_i^a = (1/2)b^{\dagger}_{i, \a}\s^a_{\a\b}b_{i, \b}\) using Table \ref{table: Noncoplanar RMOs expectations}.

Now let us show how the the action in Eq.\eqref{matrix field theory} recovers the desired phase diagram hosting a CSL and a noncoplanar ordered state. For \(g<g_c\), the partons \(U\) condense and gain a vacuum expectation value \(U_0\), breaking SO$(3)$ symmetry which fixes some reference frame in spin space for the ordering. The gauge field gains a Higgs mass in the process which gaps the Chern-Simons term. The resulting low energy field theory on this symmetry broken side reduces to  a SO$(3)$ NL\(\s\)M without a Berry phase term (see Appendix \ref{theory correspondence})
\begin{equation}
    \label{SO(3) NLSM}
    \ml_{R} = \frac{1}{2g}\Tr\left(\partial_{\m}R^T\cdot \partial_{\m}R\right) = \frac{1}{2g}\sum_i \partial_{\m}\hat{n}_i\cdot \partial_{\m}\hat{n}_i\,,
\end{equation}
where \(R^{T}\cdot R = \mathds{1}\) or equivalently \(\hat{n}_i\cdot \hat{n}_j = \d_{ij}\). The \(R\) fields condense at some mean-field \(R_0(U_0)\) such that \(\expval{n_i^a} = \left(R_0 \cdot e_i\right)^a\). The ground state clearly breaks the \(\mathrm{SO}(3)_L\times \mathrm{SO}(3)_R\) symmetry. However, it does not break it completely. The ground state is invariant under \(R_0\rightarrow \mr_L\cdot R_0 \cdot \mr_R^{T}\) where \(\mr_{L/R}\in \mathrm{SO}(3)_{L/R}\) and \(\mr_L = R_0\cdot \mr_R\cdot R_0^{T}\). This is the diagonal subgroup of \(\mathrm{SO}(3)_D^{R_0} \subset\mathrm{SO}(3)_L\times \mathrm{SO}(3)_R\) rotated by \(R_0\). The residual lattice symmetries of the ordered state, \(H=p6\), enumerated in Table \ref{table: Noncoplanar RMOs} form a discrete non-abelian subgroup \(p6 \subset \mathrm{SO}(3)_D^{R_0} \) of this continuous symmetry emergent in the long-wavelength theory \cite{Tchernyshyov2025}. 
As mentioned in Sec.~\ref{sec: RMO}, \(H_0=\tilde{p}2 \subset H\) is the residual symmetry \emph{common} to all symmetry broken states. This implies that  the elements \(h_0\in H_0 = \{T_{\ba_1}^2, T_{\ba_2}^2, C_{2, a}\}\) must be embedded trivially as the identity in \(\mathrm{SO}(3)_D^{R_0}\). This in turn implies that all elements of \(H_0\) can only be embedded in the \(\Z_2\) center of \(\mathrm{SU(2)_R}\) for the partons. This effectively restricts embeddings of lattice translation to multiples of \(\pi\) rotations in \(\mathrm{SU}(2)_R\), while lattice rotations are restricted to multiples of \(2\pi/3\) rotations in \(\mathrm{SU}(2)_R\), consistent with Eq.\eqref{translation kagom\'e U} and Eq.\eqref{rotation kagom\'e U}.\par

Since there are three broken generators we find three Goldstone modes after spontaneous symmetry breaking \(\mathrm{SO}(3)_L\times \mathrm{SO}(3)_R \rightarrow \mathrm{SO}(3)_D^{R_0}\) in the system (see Appendix \ref{theory correspondence}), consistent with noncoplanar anti-ferromagnetic orders, and reminiscent of chiral symmetry breaking in the principal chiral model used to model pions in particle physics. The emergent \(\mathrm{SO}(3)\) symmetry in the broken symmetry phase also forces all spin-wave velocities to be equal at low energies, again consistent with the octahedral or tetrahedral orders \cite{park2025}. Meanwhile, for \(g>g_c\) the partons \(U\) are gapped and can be integrated out. Hence, the low energy theory is a just a level-\(K\) U$(1)$ Chern-Simons theory which describes topological order. Particularly for \(K=2\), it describes the KL-CSL. At criticality \(g=g_c\), the theory consists of massless bosonic matrix partons coupled to a Chern-Simons gauge field. Leading order symmetry-allowed perturbations to such a critical theory include terms like \(\left|\Tr\left(U^{\dagger}\cdot\partial_{\mu}U\right)\right|^2\) which are irrelevant at tree-level. Terms which involve derivatives \emph{cannot} change the symmetry breaking order, but only renormalize the low-energy spin-wave modes. \par 

Let us reiterate that with the proper implementation of symmetries, this theory can describe such transitions between a regular noncoplanar antiferromagnetic order and a topological order on any system with \(G=p6\times \mathrm{SO}(3)\) symmetry. Examples include the aforementioned Octahedral $\leftrightarrow$ CSL on the kagom\'e lattice \cite{Bose2023} and Tetrahedral $\leftrightarrow$ CSL on the triangular lattice and honeycomb lattice \cite{Hickey2016, Hickey2017}. However, the field theory only incorporates the symmetry group and cannot distinguish between lattices. That information is given by the anomalies themselves which can appear as additional topological terms discussed in Sec. \ref{sec:anomaly}. At the level of the action, these topological terms may be implemented through coupling of the theory to background gauge fields for each symmetries in the theory.

\par 

This formalism may even be extended to systems with coplanar orderings, although the field theory in Eq.\eqref{matrix field theory} would include additional symmetry allowed terms. This is because the residual lattice symmetries (see Table \ref{table: coplanar and collinear RMOs}) are only modified by spin-rotations around the axis perpendicular to the ordering plane and are represented as an abelian subgroup \(p6 \subset \mathrm{SO}(2)_R \subset \mathrm{SO}(3)_R\). The low-energy \(\mathrm{SO}(3)_L\times \mathrm{SO}(3)_R\) symmetry would be generically lowered to \(\mathrm{SO}(3)_L\times \mathrm{SO}(2)_R\) for coplanar orders, which further undergoes spontaneous symmetry breaking down to \(\mathrm{SO}(2)_{D}\) in the ordered phase. This is completely consistent with coplanar antiferromagnetic orders reported in Table \ref{table: coplanar and collinear RMOs} which do exhibit a residual \(\mathrm{SO}(2)\) symmetry in the three low-energy spin-waves : the two modes acting in-plane with respect to the ordering plane, as opposed to one out-of-plane mode \cite{Tchernyshyov2025}. 

\section{Discussion and Outlook}
\label{sec: Discussion}

\subsection{Numerical Observables}
The  benefit of constructing such a field theory as in Eq.\eqref{eq:field-theory} and Eq.\eqref{matrix field theory} is twofold. It provides conceptual clarity about how to interpret the relevant degrees of freedom at such a transition. 
The more tangible benefit is the predictive ability of a continuum field theory. 
The \(O(3)\) NL$\s$M constructions are strongly interacting field theories which are not-renormalizable in perturbative expansion of the coupling \cite{Shnir_2018}.
However, the large-\(N\) limit of the theory where \(O(3)\rightarrow O(N)\) \cite{Hooft2002, Brezin1976, Sachdev1994, Bijnens2011} can be treated with diagrammatics where critical exponents of the theory may then be estimated, for example using $\epsilon$-expansion approximations.
This procedure can also be applied to the large-\(N\) limit of the SO\((3)\) matrix sigma-model as well, or the large-\(N\) limit of the three coupled \(\mathbb{CP}^{1}\) fields on the disordered side \cite{Eguchi1994, Rossi2016}.\par

Quantum Monte Carlo simulations of these field theories in Eq. \ref{eq:field-theory} and Eq. \ref{matrix field theory} could shed light on features of the transition, as it has for other field theories to describe transitions out of spin liquids, such as the Gross-Neveu-Yukawa universality class for transitions out of Dirac systems \cite{Boyack2021}.
While QMC simulations of frustrated spin systems suffer from sign-problems, there are sign-problem free formulations \cite{Shackleton2025}. At criticality, the relevant strictly bosonic portion of Eq.\eqref{eq:field-theory} and Eq.\eqref{matrix field theory} would also be amenable to large-scale sign-problem free QMC studies.  
Incorporating the Chern-Simons dynamics in lattice simulations
requires particular care in the framing and regularization of the problem \cite{Witten1989, Bar-natan1991}. The general simulation of $\textrm{SU}(2)_k $ Chern-Simons theory in turn suffers from an `intrinsic' sign problem \cite{Golan2020}. In fact, simulating non-abelian theories ($k \geq 2$) is equivalent to universal quantum computation and is likely out of reach of classical computations \cite{Freedman2002,  Nayak2008,Aharonov2009, Kuperberg2015}. The simpler abelian $\mathrm{U}(1)_2 \equiv \mathrm{SU}(2)_1$ theory relevant here however may  be classically simulatable. While restricted to the ordered side, there has also been considerable progress on simulations of NL$\s$M theories with topological terms \cite{Azcoiti2007, Alles2008, Azcoiti2012, Nguyen2023}. 
Further research on both analytical and numerical methods in concordance with one another could determine the universality class of these matter-gauge theories described in Sec. \ref{sec: field theory}. \par
Our work calls for additional investigations into lattice models demonstrating CSL to noncoplanar transitions. The microscopic chiral spin model with additional long range Heisenberg interaction considered in \cite{Bose2023} which realizes a possible continuous CSL to octahedral order transition on the kagom\'e lattice may warrant further investigation along with related models on other lattices \cite{ Hickey2016, Gong2017, Wietek2017, Oliviero2022}. \par

There has been progress in understanding the nature of continuous transitions out of CSLs within tensor network methods \cite{Divic2024,Kuhlenkamp2024, Divic2024b} through diverging correlation lengths and careful analysis of the bulk-boundary correspondence. 
Tensor network methods can also provide insight through analysis of entanglement entropy around the transition \cite{Cirac2021}. 
There is also a need to connect the measurements of symmetry fractionalization through tensor-network methods \cite{Zalatel2016, Zalatel2017} with the TQFT approaches \cite{Ye2024}.

\subsection{Outlook}

In this paper, we provide a topological definition of regular magnetic orders (RMOs) which takes into account their symmetries as well as their topological invariants. We then state a general condition for continuous transitions to be possible between a CSL and an RMO: the anomaly of the CSL must be compatible with the topological invariants of the RMO. We develop a class of continuum field theories to describe continuous transitions between noncoplanar spin-crystals and the KL $\mathrm{U}(1)_2$ CSL determined by compatibility relations. We construct two distinct formulation of the transition using  (a) multi-flavor \(\mathbb{CP}^1\) models coupled via a chiral interaction, and (b) a \(\mathrm{U}(2)\) principal chiral model with the transition being analogous to chiral symmetry breaking in the principal chiral model.\par
 
Once a microscopic lattice, or realization of $p6$, has been specified, these compatibility conditions are quite generous. We expect that transitions to all quantum RMOs in $p6$ with broken TRS should be allowed. Since the cuboc state has additional reflection symmetry compared to the CSL, the CSL-to-cuboc transition might need to be highly unconventional as one cannot formulate the CSL as a fluctuating cuboc-1 order \cite{Gong2015}. The cuboctahedral order also has unconventional, or `non-Lagrangian' transitions to valence-bond-solid order \cite{Zou2021}. On the other hand, the transitions from CSL to TRS broken regular noncoplanar orders have an appealing picture as the spin chirality remains broken on either side of the transition. \par

The first field theory construction makes use of the infrastructure of the NL$\s$M with a parameterization of three NL\(\s\)M theories coupled via a chiral interaction. While this formulation has intuitive appeal for the kagom\'e lattice in particular, there are a few technical difficulties which are resolved by the more general matrix parton formulation. 
We analyze the phases in the field theory and discuss emergent low energy symmetries in the ordered phase and their implications on spin-waves. 
We further discuss the generalization of the theory to possible coplanar orderings.
Although the precise solution of this class of theories remains an open research topic, this work encourages additional analytical and numerical progress that could elucidate the nature of such interesting transitions including studies of
lattice model to search for unambiguous explicit realizations of continuous transitions of this form. \par

These more complicated vector or matrix theories do not have an obvious fermionic dual, although discovering a potential dual theory to the theories discussed here remains a fascinating open possibility which would provide a complementary analysis of such a transition that may aid in solving the theory. In Appendix~\ref{app:fermion-duals}, we discuss some fermion descriptions dual to the scalar theories discussed in the main text, which we derive using the conjectured ``web of dualities" \cite{Seiberg2016, Wang2017}. While these constitute valid dual theories (to the extent that the web of dualities holds), they involve a large number of emergent gauge fields, suggesting that they may not be the most ``natural" dual descriptions. We leave the identification of such a natural dual theory to future work.
Although the bosonic NL$\s$M is a strongly coupled interacting field theory,  potential large-\(N\) approximations and more rigorous numerical approaches could be utilized in future work. 
This would provide a quantitative understanding of the theory, such as extracting critical exponents and other observables. \par

\begin{acknowledgments}
    We thank Andrew Cox, R. Ganesh, Yin-Chen He, Joseph Maciejko, Jyotirmoy Roy,  Chong Wang, Cenke Xu, Weicheng Ye, and Liujun Zou for many insightful discussions. N.M. thanks Dominic Else for a related collaboration. We acknowledge support from the Natural Sciences and Engineering Research Council (NSERC) of Canada. A.H. acknowledges support from a NSERC Graduate Fellowship (PGS-D). Research at Perimeter Institute is supported in part by the Government of Canada through the Department of Innovation, Science and Economic Development and by the Province of Ontario through the Ministry of Colleges and Universities. 
\end{acknowledgments}

\appendix

\section{Parameterized families and compatibility relations}\label{app:families}

This appendix is devoted to describing the mathematical details underlying the classification of parameterized families and the derivation of the compatibility relation. 
We first briefly review of the classification of parameterized families, following Ref.~\cite{Manjunath2024}. More details about their general definition can be found in Refs.~\cite{Wen2023}.
We will use definitions and basic results from cohomology theory without detailed derivation, but try to explain their physical meaning as clearly as possible.
Some useful background material oriented towards physicists can be found in \cite{Chen2013,manjunath2020fqh}, and prior cohomology computations for quantum spin liquids have been done in Refs.~\cite{Ye2022LSM,Ye2024}.
Many useful results on computing homotopy groups can be found in the classic review \cite{Mermin1979review}.
\subsection{Classification of parameterized families}

In the main text, we stated that a parameterized family is defined by two levels of data: the symmetry data $(G,H,H_0,\Lambda = G/H)$ and a topological invariant which depends on $\Lambda$ and $H_0$.
Below we will clarify the nature of this topological invariant. 
Note that if we were only interested in the gapped region of parameter space, it would be sufficient to only include $\Lambda$ and $H_0$ in the data for the family. 
However, it is natural to also include $G$ and $H$ in the data whenever we are also interested in the symmetry-breaking transition that produces the family.

Here we will only consider families of ordered states in which each state has trivial topological order (these are called invertible families). 
Two states in the family can always be adiabatically connected to each other, i.e. they lie in the same phase of matter. 
However the family can still have a nontrivial topological invariant, which prevents us from connecting each state to a trivial product state without closing the gap for at least one state. 
The topological invariant is classified as follows. 
For bosonic invertible families in $d$ spatial dimensions, the classification is given by the cohomology group $H^{d+2}(\Lambda \times BH_0,\Z)$ \footnote{Notation: $H$ denotes singular cohomology while $\HH$ denotes group cohomology.
The two are related as follows.
For a group $G$, it is typically true (and always true for the examples in this paper) that $H^n(BG,\Z) = \HH^n(G,\Z)$. 
It is also typically true (with some exceptions) that $\HH^n(G,\Z) = \HH^{n-1}(G,U(1))$, the latter being the group that most commonly appears in the SPT literature.}. Here $BH_0$ is the classifying space of $H_0$. Note that if $\Lambda$ is trivial, the classification reduces to that of bosonic SPT phases in $d$ dimensions, which is $H^{d+2}(BH_0,\Z)$. In fact, it is always true that
\begin{align}
    H^{d+2}(\Lambda \times BH_0,\Z) & = H^{d+2}( BH_0,\Z) \nonumber \\ & \quad\times \{H_0-\text{pumps over }\Lambda\}_d.
\end{align}

This means that the classification of all families can be split into two factors.
The first factor classifies ordinary SPT phases, i.e. families in which each state is the same SPT state and there is no dependence on the order parameter.
We generally ignore this factor as it is not relevant to the ordered states of interest.
The second factor classifies all families that depend nontrivially on the order parameter (i.e. they would be trivial if we demanded the order parameter to remain constant).
This will be the relevant classification of ordered states in our case.

Now let us discuss some examples. 

\subsubsection{0d spin in a magnetic field} 
For a zero-dimensional spin in a magnetic field, $H_0$ is trivial and $\Lambda = S^2$. 
Since $H_0$ is trivial we do not have to worry about contributions from SPT phases. 
Here we have
\begin{equation}
    H^2(S^2,\Z) = \Z,
\end{equation}
where the $\Z$ invariant is given by the Berry-Chern number.

\subsubsection{2d ferromagnet with symmetry group, $H = SO(2) \times \Gsp$}
For a ferromagnet in two space dimensions with an orientation-preserving wallpaper group symmetry $\Gsp$, we can simplify the classification result as follows, using the Kunneth formula:
\begin{align}
    & H^4(S^2 \times B\Gsp,\Z) \nonumber \\  = & H^4(B\Gsp,\Z) \times H^2(B\Gsp,H^2(S^2,\Z)) \nonumber \\ = & H^4(B\Gsp,\Z) \times  H^2(B\Gsp,\Z).
\end{align}
The first factor classifies SPT phases with $\Gsp$ symmetry. 
We ignore this term because it does not describe the ordered states of interest. 
The second factor corresponds precisely to the classification of charge responses in bosonic SPT phases with $\mathrm{U}(1) \times \Gsp$ symmetry, and has been worked out in several places, both directly within group cohomology \cite{Manjunath2021} and using the real-space constructions discussed in the main text \cite{Huang2017}. 
This factor describes the distinct ways to obtain a ferromagnet by tiling a pattern of Berry-Chern numbers over a $\Gsp$-symmetric lattice.

\subsubsection{0d classification for general $\Lambda$, and its dependence on $\pi_1(\Lambda)$ or $\pi_2(\Lambda)$}

More generally, we can consider a parameter space $\Lambda$, and the classification of zero-dimensional families over $\Lambda$ equals the cohomology group $H^2(\Lambda,\Z)$ (assuming $H_0$ is trivial). 
This cohomology group can be related to the lowest nontrivial homotopy groups of $\Lambda$, and the precise interpretation can be resolved into different cases:\footnote{Here we use the Universal Coefficient Theorem which relates cohomology and homology, together with the Hurewicz theorem which relates homology to homotopy.}
\begin{enumerate}
    \item Suppose the fundamental group $\pi_1(\Lambda)$ is trivial, i.e. all loops in parameter space are contractible, but $\pi_2(\Lambda)$ is nontrivial. Then the family classification is determined by $\pi_2(\Lambda)$:
    \begin{equation}
        H^2(\Lambda,\Z) \cong \operatorname{Hom}(\pi_2(\Lambda),\Z).
    \end{equation}
    The group $\operatorname{Hom}(A,B)$ classifies group homomorphisms from $A$ to $B$. The interpretation of the family invariant is that skyrmions of $\Lambda$ are assigned an integer, which is just the Berry-Chern number when $\Lambda = S^2$.
    \item If $\pi_1(\Lambda) \neq 0$ but $\pi_2(\Lambda) = 0$, we instead have
    \begin{equation}
        H^2(\Lambda,\Z) \cong \operatorname{Ext}(\pi_1(\Lambda)_{ab},\Z) \cong \pi_1(\Lambda)_{ab}.
    \end{equation}
    The group $\pi_1(\Lambda)_{ab}$ is called the \textit{abelianization} of $\pi_1(\Lambda)$. The abelianization of a group $G$ is defined as $G_{ab} := G/\{x y x^{-1} y^{-1} \forall x,y \in G\}$), i.e. we quotient out the group by all nontrivial commutators. $G_{ab} = G$ by definition whenever $G$ is abelian. 
    
    The group $\operatorname{Ext}(\pi_1(\Lambda)_{ab},\Z)$ classifies all possible group extensions of $\pi_1(\Lambda)_{ab}$ by $\Z$. Equivalently it classifies all possible assignments of rational numbers to elements of $\pi_1(\Lambda)_{ab}$, $\gamma \rightarrow f(\gamma) \in \mathbb{Q}$ such that the quantity 
    \begin{equation*}
        f(\gamma_1) + f(\gamma_2) - f(\gamma_1 \circ \gamma_2)
    \end{equation*}
    is always an integer.

Physically, this classification means that if the order parameter is varied along a nontrivial path $\gamma$ in $\pi_1(\Lambda)$, the ground state can acquire a Berry phase $e^{2\pi if(\gamma)}$. 
This Berry phase assignment is constrained by the condition that it should be additive over paths. 
For two paths $\gamma_1, \gamma_2$, the total Berry phase should not depend on the order in which they are traversed:
\begin{align}
    e^{2\pi if(\gamma_1 \circ \gamma_2)} &= e^{2\pi if(\gamma_2 \circ \gamma_1) } \nonumber \\\Longleftrightarrow e^{2\pi i f(\gamma_1 \circ \gamma_2 \circ \gamma_1^{-1} \circ \gamma_2^{-1}) } &= 1.
\end{align}
To implement this restriction it is natural to define $f$ from $\pi_1(\Lambda)_{ab} \rightarrow \mathbb{Q}$ as above.  
\item The most general case occurs when both $\pi_1(\Lambda)$ and $\pi_2(\Lambda)$ are nontrivial. Although it can be handled using similar techniques and reduces to a combination of the first two cases, for simplicity we will not elaborate on it here.
\end{enumerate}

\subsubsection{0d families with $\Lambda = \mathbb{RP}^3$}

Consider a zero-dimensional family over $\Lambda = \mathbb{RP}^3$, i.e. the $G = SO(3)$ symmetry is fully broken, and $H = H_0 = \Z_1$. 
Note that if we started with a single spin degree of freedom, it would not be possible to break $SO(3)$ completely, but we can do so if we start with a system of three or more spins aligned along noncoplanar axes. 

We have $\pi_1(\mathbb{RP}^3) = \Z_2$ and $\pi_2(\mathbb{RP}^3)=0$. Therefore, from point (2) above, $H^2(\mathbb{RP}^3,\Z_2) = \Z_2$. Let us first discuss what a nontrivial path in $\mathbb{RP}^3$ actually means. 
The space $\mathbb{RP}^3 \cong SO(3)$ can be visualized as a solid 3-dimensional ball of radius $\pi$, with antipodal points on the surface identified. 
The point $(\hat{n},\theta)$ corresponds to the rotation in $SO(3)$ by the angle $\theta$ about the axis $\hat{n}$. We clearly have $(\hat{n},\pi) \simeq (-\hat{n},\pi)$, which explains the above identification. 
Now a nontrivial path is any loop that connects two antipodal points on the ball.
This path can be traversed by rotating the order parameter with some sequence of rotations in $SO(3)$ that is equivalent to a $2\pi$ rotation about some axis $\hat{n}$.

The interpretation of the 0d family invariant is that if the order parameter is taken along a nontrivial path in $\mathbb{RP}^3$ (e.g. rotated by a $2\pi$ rotation in $SO(3)$), the ground state of the 0d spin system can acquire a Berry phase of either 0 or $\pi$. 

\subsubsection{2d families with $\Lambda = \mathbb{RP}^3$}

In order to classify two-dimensional families over $\mathbb{RP}^3$ with symmetry $H_0 = \tilde{p2}$, we use the above relation, together with the fact that $H^3(\mathbb{RP}^3,\Z) = \Z$ and the Kunneth formula, to evaluate
\begin{align}
    & H^4(\mathbb{RP}^3 \times BH_0,\Z) \nonumber \\ = & H^2(BH_0,H^2(\mathbb{RP}^3,\Z)) \times H^1(BH_0,H^3(\mathbb{RP}^3,\Z)) \nonumber \\
    = & H^2(BH_0,\Z_2) \nonumber \\
    = & H^2(BH_0,\Z) \otimes \Z_2.
\end{align}
If we consider just the second line, the second factor is trivial whenever $H_0$ is orientation-preserving. 
To evaluate the first factor, we perform the classification of charge responses just as in the ferromagnet (this gives $H^2(BH_0,\Z)$), but we replace all $\Z_n$ factors by $\Z_{(2,n)}$ and $\Z$ factors by $\Z_2$, where $(2,n)$ is the greatest common divisor of $2$ and $n$. 
Finally, we have $H^2(B\tilde{p2},\Z) = \Z \times \Z_2^3$ \cite{Manjunath2021}, which gives $H^4(\mathbb{RP}^3\times B\tilde{p2},\Z) = \Z_2^4$. 

\subsection{Compatibility with the anomaly}

The goal of this section is to derive a `compatibility relation' between the classification of families and that of anomalies, following Ref.~\cite{Manjunath2024}.
Such a relation exists for the following reason. Start with a parameterized family described by some topological terms in a NL$\s$M for an order parameter field valued in the space $G/H$.
We can make the order parameter field dynamical, so that the fluctuations restore the $G$ symmetry. Operationally, the order parameter field is replaced by a background $G$ gauge field.
Then we can ask how the topological action behaves under $G$ gauge transformations. 
In general, the presence of a topological term in the original NL$\s$M means that the final action is not `$G$-equivariant': the presence of the $G$ gauge field leads to phase factors in the partition function that cannot be cancelled away by any redefinitions or gauge transformations. 
This is the statement that the $G$ symmetric system has an anomaly. 
Going in reverse, we can only realize a given parameterized family by breaking $G$ symmetry if $G$ has a specific anomaly. 

The mathematical object which tells us the families that can be realized from a given $G$ anomaly, and the anomalies that could lead to a given family, is called a compatibility relation. In some cases it can be written as a function from the group of families to the group of anomalies (and the cases discussed below are of this type). 
In general however, it is not even a function, just a binary relation that needs to respect the general stacking rules for invertible states and families \cite{Manjunath2024}. 
The actual computations involves a spectral sequence (specifically the Atiyah-Hirzebruch spectral sequence) which has appeared previously in the condensed matter literature \cite{Wang2021AHSS,Shiozaki2022AHSS}; see \cite{Mccleary2000} for a readable mathematical introduction.

\subsubsection{Ferromagnet: $\Lambda = S^2$}

First let us discuss the compatibility relation for a single zero-dimensional spin in a magnetic field. Recall that the classification of ordered states (families) is given by $H^2(S^2,\Z)$. 
Now the classification of anomalies of SO$(3)$ is given by $H^3(BSO(3),\Z) = \Z_2$, with the two classes corresponding to integer spin and spin-$1/2$ representations of SO$(3)$ respectively. 
(This group also classifies SPT phases of SO$(3)$ in 1d, where the above representations characterize the projective representations carries by the edge degrees of freedom.) 
We wish to find a precise relation between these two classifications.

On physical grounds we argued that the trivial anomaly is only compatible with even Berry-Chern number, while the nontrivial anomaly is only compatible with odd Berry-Chern number. 
This can be seen more formally from the fibration
\begin{equation}
    \Lambda = S^2 \rightarrow BH = BSO(2) \rightarrow BG = BSO(3)
\end{equation}
and the associated `compatibility spectral sequence' discussed in Ref.~\cite{Manjunath2024}. 
In the present case this reduces to the Atiyah-Hirzebruch spectral sequence (AHSS)
\begin{equation}
    E_2^{p,q} = H^p(BSO(3),H^q(S^2,\Z)) \implies H^{p+q}(BSO(2),\Z).
\end{equation}
The result to note here is that this spectral sequence contains a map
\begin{equation}\label{eq:comp-FM-0d}
    H^2(S^2,\Z) \xrightarrow{\mod 2} H^3(BSO(3),\Z) 
\end{equation}
which is the same map between the Berry-Chern number of parameterized families and $G$-anomalies that we derived earlier using more physical arguments. 

Now let us consider a two-dimensional ferromagnetic state with $\Gsp = p6$ space group symmetry. 
The classification of families (after excluding the constant piece) is $H^2(p6,\Z) = \Z \times \Z_6^a$. The classification of anomalies is given by $H^5(B(SO(3) \times p6),\Z) = \Z_2^{\mathrm{LSM}} \times \Z_2^a$. The $\Z_2^{\mathrm{LSM}}$ index counts the net fractional spin per unit cell. 
By running the AHSS and identifying its differentials, we find that the analog of Eq.~\eqref{eq:comp-FM-0d} is
\begin{equation}\label{eq:comp-FM-2d}
    \Z \times \Z_6 \xrightarrow{(p,q) \rightarrow (p \mod 2, q \mod 2)} \Z_2^{\mathrm{LSM}} \times \Z_2^{a}.
\end{equation}
With our conventions, the kagom\'e lattice anomaly is specified by the $(1,0)$ element of $\Z_2 \times \Z_2$. 
The above map states that this anomaly is compatible with all the families $(p,q)$ in the $\Z \times \Z_6$ classification such that $p \equiv 1 \mod 2, q \equiv 0 \mod 2$. 
This reproduces the claim in the main text that the family is specified by an even or odd Chern number at a given high symmetry point according as the anomaly index for that high symmetry point is 0 or 1 mod 2. 

\subsubsection{noncoplanar orders with $\Lambda = \mathbb{RP}^3$}
If we take $\Lambda = \mathbb{RP}^3$ in $d=0$, the family classification is $H^2(\mathbb{RP}^3,\Z) = \Z_2$ and the $G$ anomaly classification remains $H^3(BSO(3),\Z) = \Z_2$. In this case one can use the (trivial) fibration
\begin{equation}
    \mathbb{RP}^3 \rightarrow B\Z_1 \rightarrow BSO(3)
\end{equation}
and its associated AHSS to conclude that the map 
\begin{equation}
    H^2(\mathbb{RP}^3,\Z) \rightarrow H^3(BSO(3),\Z) 
\end{equation}
is in fact an isomorphism. 
This means that a nontrivial family over $\mathbb{RP}^3$ can \textit{only} be realized if we start with a system that initially has $SO(3)$ anomaly, i.e. a net half-integer spin.

Finally, let us discuss 2d families over $\mathbb{RP}^3$ with p6 symmetry. We will specialize to the octahedral order, for which $G = SO(3) \times p6, H = p6, \Lambda = \mathbb{RP}^3, H_0 = \tilde{p2}$. After dropping the SPT piece, the classification of families is 
\begin{align}
    & H^3( B\tilde{p2},H^2(\mathbb{RP}^3,\Z ))\nonumber \\
    = & H^3(B\tilde{p2},\Z_2) = \Z_2^4. 
\end{align}
We can define the four indices as $\Z_2^{u.c.} \times \Z_2^p \times \Z_2^q \times \Z_2^r$. The first index counts the total number of $\mathbb{RP}^3$ families per unit cell of $\tilde{p2}$. 
The remaining three indices $p,q,r$ denote the three points in the $\tilde{p2}$ unit cell which correspond to kagom\'e sites in the original $p6$ unit cell. 
With this notation, the octahedral state on the kagom\'e lattice is specified by the vector $(1,1,1,1)$. 
Note that if we start with a $p6$ symmetric order, all the kagom\'e sites are either filled or unfilled, so the last three indices in the $\Z_2^4$ classification are forced to be equal.

The anomaly classification can be written as $\Z_2^{\mathrm{LSM}} \times \Z_2^a$. Similar to the previous cases, running the AHSS gives a map
\begin{equation}
    \Z_2^{u.c.} \times \Z_2^{p} \times \Z_2^q \times \Z_2^r  \xrightarrow{(m,n,n,n) \rightarrow (m,m+n)} \Z_2^{\mathrm{LSM}} \times \Z_2^a.
\end{equation}
Again, the compatibility relation is ultimately just a constraint on which sites in the real-space unit cell are allowed to host ordered degrees of freedom that transform as half-integer spin objects under $SO(3)$.

\subsection{Field-theoretic description of CSL anomaly}\label{app:anomalyterms}

In this section we will review how to describe the CSL anomaly using gauge fields for an effective internal symmetry which is isomorphic to $SO(3) \times \Gsp$, where $\Gsp = p6$. We will define the theory on a (2+1)-dimensional space-time manifold $M$ with a triangulation. The gauge fields are 1-cochains defined on the edges of the triangulation. We first define the $SO(3)$ gauge field $B$, as well as the $C_6$ rotation and translation gauge fields
\begin{align}
    \omega &\in \frac{2\pi}{6} \Z; \quad \omega \simeq \omega + 2\pi \\
    \vec{R} = (X,Y) & \in 2\pi \Z \times 2\pi \Z. 
\end{align}
Note that the definition of $\omega$ is with respect to the $C_6$ rotation center in the unit cell. We impose the \textit{flatness} condition that the product of group elements around any 2-simplex is trivial. This leads to a constraint on the allowed holonomies and gauge transformations of $\omega$ and $\vec{R}$. See \cite{Manjunath2021} for a detailed treatment.

Flux of the rotational symmetry is given by the 2-cochain $d\omega$. We can also define an `area form' $A_{XY}$ whose integral over a spatial slice of the manifold counts the number of unit cells, and which reduces to $X \wedge Y$ when $\omega = 0$. For general choices of $\Gsp$ we can also consider non-trivially quantized dislocation fluxes \cite{Manjunath2021}, but this is not necessary for the group $p6$.

The anomaly of a 0-dimensional system with spin 1/2 is described by a topological term involving only $B$:
\begin{equation}
    \mathcal{L}_{\text{anom, 0d}} = \pi w_2[B]
\end{equation}
where $w_2$ is the second Stiefel-Whitney class of the gauge bundle. The anomaly action is obtained by taking its cup product with each crystalline `flux':
\begin{equation}
    \mathcal{L}_{\text{anom, 0d}} = \pi w_2[B] \cup \left(\frac{k_1}{2\pi} d\omega + \frac{k_2}{2\pi} A_{XY} \right)
\end{equation}
where $k_1, k_2 \in \Z_2$ are the two anomaly invariants. The term with $k_1=1$ assigns a net spin-1/2 to each unit cell, i.e. it is the action for the usual LSM anomaly. The term with $k_2 = 1$ assigns a net spin-1/2 at each 6-fold rotation center ($a$ site) in the unit cell. 

Instead of working with translation gauge fields and area forms, we could introduce gauge fields $\omega_o, o = a,b,c$ for the $C_6, C_3$ and $C_2$ rotation centers $a,b,c$. In this case, the above anomaly could be re-expressed as
\begin{equation}
    \mathcal{L}_{\text{anom, 0d}} = \pi w_2[B] \cup \left(\frac{k_1+k_2}{2\pi} d\omega_c + \frac{k_2}{2\pi} d\omega_a \right).
\end{equation}
We can switch between representations based on convenience.

\begin{table*}[!ht]
\centering
\begin{tabular}{ |c|c|c|c|c|c|c|c| } 
     \hline
     Order & Lattice & \(\tilde{T}_{\ba_1}\) & \(\tilde{T}_{\ba_2}\) & \(\tilde{C}_{3, a}\) & \(\tilde{C}_{3, b}\) & \(\tilde{C}_{2, a}\) & \(\tilde{C}_{2, c}\)\\
     \hline 
     Collinear N\'eel & Honeycomb & \(\mathds{1}\times T_{\ba_1}\) & \(\mathds{1}\times T_{\ba_2}\) & \(\mathds{1}\times C_{3, a}\) & \(\mathds{1}\times C_{3, b}\) & \(\mr_{\pi}(\hat{x})\times C_{2, a}\) & \(\mr_{\pi}(\hat{x})\times C_{2, c}\) \\
     \hline 
     Coplanar \(\bQ = \boldsymbol{\Gamma}\) & kagom\'e & \(\mathds{1}\times T_{\ba_1}\) & \(\mathds{1}\times T_{\ba_2}\) & \(\mr_{2\pi/3}(\hat{z})\times C_{3, a}\) & \(\mr_{2\pi/3}(\hat{z})\times C_{3, b}\) & \(\mathds{1}\times C_{2, a}\) & \(\mathds{1}\times C_{2, c}\) \\
     \hline 
     Coplanar \(\bQ = \pm \bK\) & Triangular & \(\mr_{2\pi/3}(\hat{z})\times T_{\ba_1}\) & \(\mr_{-2\pi/3}(\hat{z})\times T_{\ba_2}\) & \(\mathds{1}\times C_{3, a}\) & \(\mr_{2\pi/3}(\hat{z})\times C_{3, b}\) & \(\mr_{\pi}(\hat{x})\times C_{2, a}\) & \(\mr_{\pi}(\hat{m})\times C_{2, c}\) \\
     \hline 
     Coplanar \(\bQ = \pm \bK\) & kagom\'e & \(\mr_{2\pi/3}(\hat{z})\times T_{\ba_1}\) & \(\mr_{-2\pi/3}(\hat{z})\times T_{\ba_2}\) & \(\mathds{1}\times C_{3, a}\) & \(\mr_{2\pi/3}(\hat{z})\times C_{3, b}\) & \(\mr_{\pi}(\hat{x})\times C_{2, a}\) & \(\mr_{\pi}(\hat{m})\times C_{2, c}\) \\
     \hline 
\end{tabular}
\caption{The modified residual lattice symmetries of all regular anti-ferromagnetic coplanar and collinear orders in systems with \(G=p6\times \mathrm{SO}(3)\) symmetry where \(p6\) includes translations \(T_{\ba_1}\,,T_{\ba_2}\) and rotations \(C_{6, a}\). These orders either have the same unit cell as the lattice implying their characteristic wave-vector is  \(\bQ=\boldsymbol{\Gamma}\), or can have an enlarged \(\sqrt{3}\times \sqrt{3}\) unit cell implying \(\bQ = \pm \bK\). Each modified lattice symmetry has an attached spin rotation \(\mr_{\theta}(\hat{n})\) by an angle \(\theta\) around a direction \(\hat{n}\) on the Bloch sphere. \(\hat{m}=-(1/2)\hat{x}-(\sqrt{3}/2)\hat{y}\).}
\label{table: coplanar and collinear RMOs}
\end{table*}

\begin{table*}[!ht]
\centering
\begin{tabular}{ |c|c|c|c|c|c|c|c| } 
     \hline
     Order & Lattice & \(\tilde{T}_{\ba_1}\) & \(\tilde{T}_{\ba_2}\) & \(\tilde{C}_{3, a}\) & \(\tilde{C}_{3, b}\) & \(\tilde{C}_{2, a}\) & \(\tilde{C}_{2, c}\)\\
     \hline 
     Collinear Ferromagnet & Honeycomb & \(\mathds{1}\times T_{\ba_1}\) & \(\mathds{1}\times T_{\ba_2}\) & \(\mathds{1}\times C_{3, a}\) & \(\mathds{1}\times C_{3, b}\) & \(\mathds{1}\times C_{2, a}\) & \(\mathds{1}\times C_{2, c}\) \\
     \hline 
     Collinear Ferromagnet & Triangular & \(\mathds{1}\times T_{\ba_1}\) & \(\mathds{1}\times T_{\ba_2}\) & \(\mathds{1}\times C_{3, a}\) & \(\mathds{1}\times C_{3, b}\) & \(\mathds{1}\times C_{2, a}\) & \(\mathds{1}\times C_{2, c}\)\\
     \hline 
     Collinear Ferromagnet & kagom\'e & \(\mathds{1}\times T_{\ba_1}\) & \(\mathds{1}\times T_{\ba_2}\) & \(\mathds{1}\times C_{3, a}\) & \(\mathds{1}\times C_{3, b}\) & \(\mathds{1}\times C_{2, a}\) & \(\mathds{1}\times C_{2, c}\)\\
     \hline 
     Umbrella & kagom\'e & \(\mathds{1}\times T_{\ba_1}\) & \(\mathds{1}\times T_{\ba_2}\) & \(\mr_{2\pi/3}(\hat{n})\times C_{3, a}\) & \(\mr_{2\pi/3}(\hat{n})\times C_{3, b}\) & \(\mathds{1}\times C_{2, a}\) & \(\mathds{1}\times C_{2, c}\) \\
     \hline 
\end{tabular}
\caption{The modified residual lattice symmetries of all regular ferromagnetic orders in systems with \(G=p6\times \mathrm{SO}(3)\) symmetry where \(p6\) includes translations \(T_{\ba_1}\,,T_{\ba_2}\) and rotations \(C_{6, a}\). All these order have the same unit cell as the lattice implying their characteristic wave-vector is  \(\bQ=\boldsymbol{\Gamma}\). Each modified lattice symmetry has an attached spin rotation \(\mr_{\theta}(\hat{n})\) by an angle \(\theta\) around a direction \(\hat{n}\) on the Bloch sphere. \(\hat{n} = (1/\sqrt{3})(1, 1, 1)\).}
\label{table: ferro collinear RMOs}
\end{table*}

\section{\(\mathrm{U}(2)\) parton to \(\mathrm{SO}(3)\) principal chiral model }\label{theory correspondence}
In this section we shall show how the \(\mathrm{U}(2)\) matrix parton theory reduces to the expected \(\mathrm{SO(3)}\) principal chiral model on the ordered side where the \(\mathrm{U}(1)\) photon are Higgsed and consequently gapped at low energies. 
Our goal is to show that the Lagrangian \(\ml_{U}\) in Eq.\eqref{matrix field theory} reduces to the Lagrangian \(\ml_R\) in Eq.\eqref{SO(3) NLSM} when integrating out the gapped photon \(a_{\m}\).

\subsection{\(\mathrm{SO}(3)\) NL\(\s\)M}
Using Eq.\eqref{matrix partons}, we want to express \(\ml_R = \Tr\left(\partial_{\m}R^{T}\cdot \partial_{\m}R\right)\) in terms of \(U\). Using the shorthand \(v_i = U\cdot u_i\) such that \(v_i^{\dagger}\cdot v_i = 1\), we get
\begin{equation}
    \label{R derivative using v}
    \partial_{\m}R_{ai} = \partial_{\m}v_i^{\dagger}\cdot \s^a \cdot v_i + v_i^{\dagger}\cdot \s^a \cdot \partial_{\m}v_i\,.
\end{equation}
Then we use the completeness relation of the Pauli matrices \(\s^a_{\a\b}\s^a_{\g\d} = 2\d_{\a\d}\d_{\b\g}-\d_{\a\b}\d_{\g\d}\), we get
\begin{equation}
    \label{R Lagrangian using U}
    \ml_{R} = \frac{1}{2g}\left(2 \ml_{U}^{(a)}-\ml_{U}^{(b)}\right)\,,
\end{equation}
where
\begin{equation}
    \label{L using U}
    \begin{split}
      \ml_{U}^{(a)} &= \left(\partial_{\m}v_i^{\dagger}\cdot v_i\right)^2+\left(v_i^{\dagger}\cdot \partial_{\m}v_i\right)^2 + 2\left(\partial_{\m}v_i^{\dagger}\cdot \partial_{\m}v_i\right)\,,\\
      \ml_{U}^{(b)} &= \left(\partial_{\m}v_i^{\dagger}\cdot v_i\right)^2+\left(v_i^{\dagger}\cdot \partial_{\m}v_i\right)^2 + 2\left|\partial_{\m}v_i^{\dagger}\cdot v_i\right|^2\,.
    \end{split}
\end{equation}
This allows us to simplify the Lagrangian as
\begin{equation}
    \label{R Lagrangian using U simplified}
    \ml_{R} = \left(\partial_{\m}v_i^{\dagger}\cdot v_i - v_i^{\dagger}\cdot \partial_{\m}v_i\right)^2 + 4\left(\partial_{\m}v_i^{\dagger}\cdot \partial_{\m}v_i\right)\,.
\end{equation}
The last step is to express Eq.\eqref{R Lagrangian using U simplified} purely in terms of \(U\). 
\begin{equation}
    \label{trace 1}
    \begin{split}
        \left(\partial_{\m}v_i^{\dagger}\cdot \partial_{\m}v_i\right) &= u_i^{\dagger}\cdot \partial_{\m}U^{\dagger} \cdot \partial_{\m}U \cdot u_i\\
        & = \sum_{i=1}^{3}\Tr\left(\partial_{\m}U^{\dagger} \cdot \partial_{\m}U \cdot E_i\right)\,,\:\:\:E_i = e_i\cdot e_i^{\dagger}\,,
    \end{split}
\end{equation}
where using the reference spinors discussed in the main text, \(E_i = (1/2)\mathds{1}+(1/2)\s^i\). Therefore, Eq.\eqref{trace 1} simplifies to 
\begin{equation}
    \label{trace 1 simplified}
    \begin{split}
        \left(\partial_{\m}v_i^{\dagger}\cdot \partial_{\m}v_i\right) &= \frac{3}{2}\Tr\left(\partial_{\m}U^{\dagger} \cdot \partial_{\m}U\right)+\frac{1}{2}\sum_{i=1}^{3}\Tr\left(\partial_{\m}U^{\dagger} \cdot \partial_{\m}U \cdot \s^i\right)\,.
    \end{split}
\end{equation}
Next, we tackle the first term in Eq.\eqref{R Lagrangian using U simplified}. Using the same trick as above, we can rewrite it as a trace as 
\begin{equation}
    \label{trace 2}
    \begin{split}
     & \left(e_i^{\dagger}\cdot M \cdot e_i\right)^2 = \Tr\left(M \cdot E_i \cdot M \cdot E_i\right)\\
     & = \sum_{i=1}^{3}\left\{\frac{1}{4}\Tr\left(M^2\right)+\frac{1}{2}\Tr\left(M^2 \s^i\right)+\frac{1}{4}\Tr\left((M \cdot \s^i \cdot M \cdot \s^i\right)\right\}\\
     & = \frac{3}{4}\Tr\left(M^2\right) + \frac{1}{2}\sum_{i=1}^{3}\Tr\left(M^2 \s^i\right) +\frac{1}{4}\Tr\left(M^2\right)-\frac{1}{4}\left(\Tr\left(M\right)\right)^2\\
     & =  \Tr\left(M^2\right) + \frac{1}{2}\sum_{i=1}^{3}\Tr\left(M^2 \s^i\right) -\frac{1}{4}\left(\Tr\left(M\right)\right)^2\,.
    \end{split}
\end{equation}
Substituting \(M=\partial_{\m}U^{\dagger}\cdot U - U^{\dagger}\cdot \partial_{\m}U\), we find that 
\begin{equation}
    \label{M^2}
    M^2 = \partial_{\m}U^{\dagger}\cdot U \cdot \partial_{\m}U^{\dagger}\cdot U + U^{\dagger}\cdot \partial_{\m}U\cdot U^{\dagger}\cdot \partial_{\m}U - 2 \partial_{\m}U^{\dagger}\cdot \partial_{\m}U\,,
\end{equation}
such that the trace simplifies to (using \(\partial_{\m}U\cdot U^{\dagger} = -U \cdot \partial_{\m}U^{\dagger}\))
\begin{equation}
    \label{M^2 trace}
    \begin{split}
     \Tr\left(M^2\right) &= - 4 \Tr \left( \partial_{\m}U^{\dagger}\cdot \partial_{\m}U\right)\,,\\
     \Tr\left(M^2 \cdot \s^i\right) &= - 4 \Tr \left( \partial_{\m}U^{\dagger}\cdot \partial_{\m}U \cdot\s^i\right)\,.
    \end{split}
\end{equation}
Therefore, we finally arrive at the \(\mathrm{SO}(3)\) NL\(\s\)M written in terms of the \(U\) fields as 
\begin{equation}
    \label{So(3) NLSm U}
    \ml_R = \frac{1}{g}\Tr\left(\partial_{\m}U^{\dagger}\cdot \partial_{\m}U\right)-\frac{1}{8g}\left(\Tr\left(U^{\dagger}\cdot \partial_{\m}U-\partial_{\m}U^{\dagger}\cdot U\right)\right)^2\,.
\end{equation}

\subsection{Integrating out the photon}
We integrate out the photon in Eq.\eqref{matrix field theory} and show that it indeed matches Eq.\eqref{So(3) NLSm U}. 
This is equivalent to solving the equation of motion for \(a_{\m}\) since the gauge field only appears quadratically in the Lagrangian. 
We also neglect the Chern-Simons term since the photon is assumed to be gapped in the symmetry broken phase. 
The equation of motion is as follows
\begin{equation}
    \label{EOM of photon}
    a_{\m} = \frac{i}{4}\Tr\left(U^{\dagger}\cdot \partial_{\m}U-\partial_{\m}U^{\dagger}\cdot U\right)\,.
\end{equation}
Substituting Eq.\eqref{EOM of photon} back into E.\eqref{matrix field theory}, we find the same Lagrangian as found in Eq.\eqref{So(3) NLSm U}. Therefore, we find that
\begin{equation}
    \label{Lagrangian matching}
    \ml_{U} \xrightarrow{\mathrm{Symmetry \: Breaking}} \ml_R\,,
\end{equation}
the \(\mathrm{U}(2)\) matrix parton theory coupled to a \(\mathrm{U}(1)\) gauge field reduces to \(\mathrm{SO}(3)\) matrix nonlinear sigma-model upon \(\mathrm{SO}(3)\) symmetry breaking.

\subsection{Goldstone modes in the theory}
Now that we have shown that the matrix parton theory reduces to the \(\mathrm{SO}(3)\) NL\(\s\)M model once the photon become Higgsed and can be integrated out. Let us now study the low-energy theory in the symmetry broken phase. We expand the field \(R\) around its mean-field value \(R_0\) as 
\begin{equation}
    \label{mean-field excitations}
    \begin{split}
     R(x) &= \exp\left(\phi_a(x) T^a\right)\cdot R_0 \cdot \exp\left(-\psi_b(x) T^b\right) \\
     & \approx R_0 + \phi_a(x) T^a\cdot R_0 - \psi_a(x) R_0\cdot T^a + \mo(\phi^2\,,\psi^2)\,,    
    \end{split}
\end{equation}
where \(T^a\) are the real, traceless, and anti-symmetric generators of \(\mathrm{so}(3)\) algebra. Expanding Eq.\eqref{SO(3) NLSM} in terms of Eq.\eqref{mean-field excitations} we find (upto a constant)
\begin{equation}
    \label{symmetry broken theory}
    \begin{split}
        \ml = \frac{1}{g} \partial_{\m}\Psi \cdot M \cdot \partial_{\m}\Psi + ...\,,\:\:\:\Psi = (\phi_a\,,\psi_a)\,,
    \end{split}
\end{equation}
where we have used trace identity \(\Tr\left(T^a\cdot T^b\right)=-2\d_{ab}\) for the generators and
\begin{equation}
    \label{M matrix}
    M_{6\times 6} = 
    \begin{pmatrix}
        \mathds{1}_{3\times 3} & -D_0\\
        -D_0^{T} & \mathds{1}_{3\times 3}
    \end{pmatrix}\,,\:\:\:D_0\in \mathrm{SO}(3)\,.
\end{equation}
The rotation matrix \(D_0\) is fixed from the adjoint representation of the generators, \(R_0^{T}\cdot T^a \cdot R_0 = D_0(R_0)^{ab}T^b\). 
One can show that \(M\) has eigenvalues \(\{2, 2, 2, 0, 0, 0\}\), the nonzero value corresponds to the three Goldstone modes \(\Psi_G = (\phi, -D_0^{T}\cdot \phi)\), while the three vanishing values correspond to the generators of the residual \(\mathrm{SO}(3)_{D}^{R_0}\) symmetry \(\Psi_S=(\phi, D_0^{T}\cdot\phi)\). The low-energy theory in the symmetry broken phase takes the form of 
\begin{equation}
    \label{symmetry broken theory 2}
    \begin{split}
        \ml = \frac{2}{g} \left(\partial_{\m}\phi\right)^2 + \mo(\phi^4)\,.
    \end{split}
\end{equation}
Now, the action of the residual symmetry is found by considering the Goldstone modes in Eq.\eqref{mean-field excitations} with \(\psi = -D_0^{T}\cdot \phi\) which gives \(R \approx R_0+2\phi_a T^a\cdot R_0\). Then under a symmetry transformation \(R \rightarrow \mr_R\cdot R \cdot \mr_R^T\) where \(\mr_L = R_0^T\cdot \mr_R \cdot R_0\), we find \(R \rightarrow R_0 + \phi_a \mr_L \cdot T^a \cdot \mr_L^{T} \cdot R_0\) which is equivalent to \(\phi \rightarrow \mr_L\cdot \phi\) where \(\mr_L\in \mathrm{SO}(3)\). 
The spin-waves  act as vectors under the residual \(\mathrm{SO}(3)_D^{R_0}\) symmetry. 
As a result, the spin-wave velocities are forced to be equal for all three modes.

\section{ Dual Fermionic Descriptions of the Transition} \label{app:fermion-duals} 

Recent years have seen the proposal of a ``web of dualities'' relating critical theories of scalars or fermions coupled to Chern-Simons gauge fields \cite{Seiberg2016, Wang2017}. Dualities between these Chern-Simons-matter theories may be viewed, in the case where they relate a fermionic theory to a bosonic theory, as relativistic $2+1d$ versions of bosonization (or, fermionization, in the reverse direction) in which, for instance, a scalar in the original theory corresponds to a composite of a fermion and a monopole in the dual theory. 
While these dualities typically relate strongly interacting theories to other strongly interacting theories, and so need not immediately lead to a solution of the original theory in question, they  have 
found manifold applications in the context of, among others, the physics of fractional quantum Hall states, spin liquids, and beyond-Landau criticality by relating seemingly distinct physical problems \cite{Senthil2019}.


With this in mind, in this Appendix, we discuss potential dual fermionic descriptions of the spinon theory, Eq.~\eqref{eq:field-theory}. There are, in principle, many such dual fermionic theories. For our purposes, 
a dual theory would be ``useful" if, for instance, the global symmetries have a simple action on the fundamental fields of the theory, in contrast to, for instance, the spinon theory of Eq.~\eqref{eq:field-theory}, in which translation permutes the gauge fields in addition to acting on the scalars. Such a dual theory may provide insight into how the bosonic descriptions of the transition discussed in the main text relate to fermionic parton theories, the latter of which may allow for better understanding the transition analytically (assuming, that it is, in fact, continuous).

Unfortunately, as we discuss below, we are unable to find a dual theory satisfying these desiderata. One major obstacle is that the web of dualities typically relates theories only possessing global $U(1)$ symmetries. 
In the present context, the theories of interest, being spin systems, have global $SU(2)$ symmetry. Progress in the development of dualities relating theories with explicit global $SU(2)$ symmetries and $U(1)$ gauge fields is more limited---such a duality relating the $SU(2)$ symmetric NCCP$^{1}$ model to a so-called $\mathrm{QED}_3$-Gross-Neveu theory was proposed in Ref.~\cite{Wang2017}, which goes beyond the standard duality web. 
Motivating and finding evidence for an analogous $SU(2)$-symmetric duality for the multi-component spinor theory of Eq.~\eqref{eq:field-theory} is a challenging task. Here, we set ourselves a more manageable goal and allow ourselves to break the global $SU(2)$ symmetry down to $U(1)$ and see how much mileage we can obtain from the standard web of dualities.
The resulting dual theory below thus does not describe the transitions discussed in the main text, though it may still be of use for related transitions between topological and conventionally ordered phases. 


Let us consider the spinon theory of Eq.~\eqref{eq:field-theory}, which we reproduce here in a slightly modified form:
\begin{equation}
        \ml = \frac{1}{g}\sum_{i,a}\left|D_{\m}^{i} z_{i,a}\right|^2 - |z_{i,a}|^4 + \sum_{i,j} \frac{K_{ij}}{4\pi} a^id a^j + \cdots\,
     \label{eq:spinor-theory-app}
\end{equation}
with,
\begin{align}
	K = -\begin{pmatrix}
	0 & 1 & 1 \\
	1 & 0 & 1 \\
	1 & 1 & 0
	\end{pmatrix} .
\end{align}
Here we have explicitly written out the two components of each scalar, as indexed by $a=1,2$, and made use of the shorthand $adb \equiv \e^{\a\b\g} a_{\a} \partial_{\b} b_{\g}$.
In this Lagrangian, we have tuned the spinor fields to criticality, which we indicate with the shorthand in which we drop their mass terms and include the quartic potentials $|z_{i,a}|^4$. 
These potentials \emph{explicitly} break the global $SU(2)$ spin-rotation symmetry down to a $U(1)$ symmetry.
Since we have tuned to criticality, we have also dropped the chiral interaction, which is irrelevant at this point.

In order to derive a dual fermionic description of Eq.~\eqref{eq:spinor-theory-app}, we make use of a duality which provides the fermionization of a Wilson-Fisher boson \cite{Seiberg2016,Karch2016}:
\begin{align}
	|D_A \phi|^2 - |\phi|^4 &\longleftrightarrow i \bar{\psi} \slashed{D}_a \psi - \frac{1}{2} \frac{1}{4\pi} ada - \frac{1}{2\pi} a dA - \frac{1}{4\pi} AdA \, . \label{eq:fermionization-app}
\end{align}
On the left of the arrow is the Lagrangian for a Wilson-Fisher boson, $\phi$, coupled to a background $U(1)$ gauge field $A$. 
On the right of the arrow is the Lagrangian for a single Dirac fermion, $\psi$, coupled to a $U(1)_{-1/2}$ Chern-Simons theory. Here, the level $-1/2$ Chern-Simons term should be understood as coming from integrating out a massive, regulator fermion. The arrow indicates that these theories are dual. A simple sanity check on this duality comes from noting that both theories describe a superfluid-Mott transition, as can be seen by adding and tuning mass terms on either side of the duality. Now, applying time-reversal to this duality and making use of the fact that the Wilson-Fisher theory is time-reversal invariant, we also have the duality,
\begin{align}
    |D_A \phi|^2 - |\phi|^4 &\longleftrightarrow i \bar{\psi} \slashed{D}_a \psi + \frac{1}{2} \frac{1}{4\pi} ada + \frac{1}{2\pi} a dA + \frac{1}{4\pi} AdA  \, ,\label{eq:fermionization-TR-app}
\end{align}
which differs from Eq.~\eqref{eq:fermionization-app} in the signs of the Chern-Simons levels.

The basic conceit of the duality web is that one case use a ``seed" duality, like Eq.~\eqref{eq:fermionization-app}, to generate new dualities.
Indeed, returning to Eq.~\eqref{eq:spinor-theory-app}, we can apply the above dualities to \emph{each} species of scalar. Explicitly, we apply Eq.~\eqref{eq:fermionization-app} to the scalars $z_{i,1}$ and Eq.~\eqref{eq:fermionization-TR-app} to the scalars $z_{i,2}$, treating the gauge fields $a_i$ as spectators (i.e. they play the role of $A$ in each application of the duality). The Chern-Simons terms for $a_i$ induced by the dualities cancel out, such that we obtain the dual theory of Eq.~\eqref{eq:spinor-theory-app}, 
\begin{align}
    \begin{gathered}
    \mathcal{L}' = \sum_{i,a} \left( i\bar{\psi}_{i,a} \slashed{D}_{b_{i,a}}\psi_{i,a} + (-1)^a\frac{1}{2}\frac{1}{4\pi}b_{i,a}db_{i,a} \right) \\
    + \frac{1}{2\pi}\sum_i a_i d(b_{i,1} - b_{i,2}) - \frac{1}{2\pi} \sum_{i<j} a_i d a_j \, .
    \end{gathered} \label{eq:dual-fermion-theory-app}
\end{align}
Here, the $\psi_{i,a}$ and $b_{i,a}$ are the Dirac fermions and emergent gauge fields obtained from dualizing $z_{i,a}$. 
While this is a putative fermionic dual of Eq.~\eqref{eq:spinor-theory-app}, it involves six Dirac fermions and a total of nine emergent gauge fields. 
It is difficult to motivate the presence of such a large number of gauge fields from a microscopic partonic description (we find that at most one gauge field can be integrated out while preserving the Dirac flux quantization condition). The presence of these gauge fields also obfuscate the action of global symmetries which will act on these additional fields. Thus, while Eq.~\eqref{eq:dual-fermion-theory-app} is \emph{a} dual theory for Eq.~\eqref{eq:spinor-theory-app}, it does not appear to be particularly useful.

Our other attempts at using the duality web also suffer from this issue of generating a large number of emergent gauge fields. We suspect one source of this difficulty comes from reducing the global $SU(2)$ down to $U(1)$; it is possible that there is a more natural duality for the full $SU(2)$-symmetry spinor theory which, by preserving the global symmetry, constraints the number of emergent fields in the fermionic description. Separately, the chiral interaction serves to energetically lock the spinors relative to one another. By dropping it at the critical point, we have effectively increased the number of dynamical modes, which appears to manifest in the proliferation of the number of emergent fields in the dual theory. Taking into account these two factors in the dualization does not appear to be straightforward. The derivation of dual theories to $SU(2)$-symmetry multi-component spinor theories thus remains a challenging and interesting task, which we leave to future work.


\newpage

\bibliography{apssamp}

\end{document}